\documentclass{IEEEcsmag}

\usepackage{amsmath,amssymb,amsfonts}
\usepackage{algorithmic}
\usepackage{textcomp}
\usepackage[table,xcdraw]{xcolor}
\usepackage[nolist]{acronym}
\usepackage{xurl}
\usepackage{paralist}
\usepackage{multirow}
\usepackage{listings}
\usepackage{tabularx}
\usepackage{rotating}
\usepackage{caption}
\usepackage{subcaption}
\usepackage{graphicx}
\usepackage{hhline}
\usepackage{arydshln}
\usepackage{tikz,float}
\usepackage{array}
\usepackage{xparse}
\usepackage[most]{tcolorbox}
\usepackage{xcolor}
\usepackage{xspace}
\usepackage{pgfplots, pgfplotstable}
\usepackage{siunitx,  
            booktabs, 
            caption}  
\usepackage{xcolor}
\usepackage[switch]{lineno}
\usepackage{comment}
\usepackage{cite}
\usepackage{adjustbox}
\usepackage{pdflscape}
\usepackage{tikz}

\colorlet{NextBlue}{red!25!green!50!blue!75}

\tikzset{
  font={\fontsize{10pt}{12}\selectfont}}
\usetikzlibrary{trees}
\usetikzlibrary{patterns, backgrounds}

\definecolor{ao(english)}{rgb}{0.0, 0.5, 0.0}
\definecolor{bananayellow}{rgb}{1.0, 0.88, 0.21}
\definecolor{bostonuniversityred}{rgb}{0.8, 0.0, 0.0}
\definecolor{brandeisblue}{rgb}{0.0, 0.44, 1.0}

\usepackage{forest}
\tikzset{
  treenode/.style={
      shape=rectangle,
      rounded corners,
      ultra thick,
      draw=brandeisblue,
      align=left,
      fill=white
  }
}

\usepackage[colorlinks,urlcolor=blue,linkcolor=blue,citecolor=blue]{hyperref}
\expandafter\def\expandafter\UrlBreaks\expandafter{\UrlBreaks\do\/\do\*\do\-\do\~\do\'\do\"\do\-}
\usepackage{upmath,color}

\begin{document}


\title{Journey to the Center of Software Supply Chain Attacks}

\author{Piergiorgio Ladisa}
\affil{SAP Security Research, Université de Rennes 1}

\author{Serena Elisa Ponta}
\affil{SAP Security Research}

\author{Antonino Sabetta}
\affil{SAP Security Research}

\author{Matias Martinez}
\affil{Universitat Politècnica de Catalunya-BarcelonaTech}

\author{Olivier Barais}
\affil{Université de Rennes 1, Inria, IRISA}


	



\begin{abstract}
  This work discusses open-source software supply chain attacks and proposes a general taxonomy describing how attackers conduct them. We then provide a list of safeguards to mitigate such attacks. We present our tool "Risk Explorer for Software Supply Chains" to explore such information and we discuss its industrial use-cases. 
\end{abstract}

\maketitle






\ac{OSS} is ubiquitous in modern applications. It may
constitute more than 90\% of the code of a commercial application and it is widely used
across the technology stack and the development and operation
lifecycle. Due to the complexity of the modern software supply chain, attackers
have multiple opportunities to inject malicious code into open-source components
and infect downstream users. 


In recent years, we have observed an exceptional increase in the number and type
of attacks on \ac{OSS}~\cite{sonatypeAnnualState}. For example, a recent case
affected a nightly build of PyTorch, a popular framework to build
machine-learning models: the attackers sneaked malicious code through a
dependency by abusing the dependency resolution mechanism of
\texttt{pip}~\cite{pytorchPyTorch}. Other software supply chain attacks (e.g.,
infection of SolarWind's Orion platform~\cite{9382367}) impacted suppliers of
government agencies and critical infrastructures. For this reason, several
national security agencies reported software supply chain attacks as one of the
primary threats~\cite{ENISAThreat, executiveorder} and different efforts (both
public and private) arose to increase the security of software supply chains
(e.g., SLSA\footnote{\url{https://slsa.dev/}}).

There is no doubt about the fact that the software supply chain is more and more
often under attack, and that this problem is receiving considerable attention
both by the industry and the academic community. However, we have also observed
that the existing literature is somewhat fragmented, also due to the lack of a
general, technology-independent description of how attackers inject malicious
code into \ac{OSS} projects.

In~\cite{ladisa2022taxonomy} we presented a taxonomy of attack vectors \ac{OSS}
supply chains, which are independent of specific programming languages and
technologies. Moreover, we introduced a set of general safeguards addressing the
identified attack vectors. To ease the visualization and exploration of the
taxonomy, as well as to enable its extension by the open-source and security
communities, we have developed a visualisation tool named \textit{Risk Explorer
for Software Supply Chains}~\cite{10.1145/3560835.3564546}, which is accessible
online and is released as open source.

In this work we will guide you through a journey to the center 
of \ac{OSS} supply chain attacks. 

We prepare the descent by describing the attack surface of \ac{OSS} development
model, associated risks, and the attacker model.
Then, we present the taxonomy covering \textbf{117 unique attack vectors}
related to \ac{OSS} supply chains and \textbf{33 safeguards} geared towards the
proposed taxonomy. This taxonomy, which is an 
updated version of the one presented in~\cite{ladisa2022taxonomy},
was built by examining \textbf{370 resources} encompassing
real-world attacks and scientific and grey literature. 
We dig deeper by analysing the vectors of real world attacks covered by those resources
and we contrast the prevalence of each such vector with the attention that it received
from the research community.


The journey continues with the review of the \textit{Risk Explorer for Software
Supply Chains}~\cite{10.1145/3560835.3564546} and four industrial use
cases.

We then position our taxonomy with respect to the existing frameworks
for software supply chain security.


The journey's end provides insights on the content of malicious packages and on the current
challenges that the software industry faces to secure the \ac{OSS} supply chain.

\section{Preparing the descent}

To understand the existing attack vectors, first we describe the attack surface
of the OSS supply chain, i.e., what are the \textit{systems} and
\textit{stakeholders} involved in the creation of \ac{OSS} artifacts. Then we
provide a general overview of the risks of \ac{OSS} supply chains. Finally, we
present what are the characteristics of the attacker in our context.

\subsection{Attack Surface: OSS Development Model}



\begin{figure}[htp]
    \centering
    \includegraphics[width=.5\textwidth]{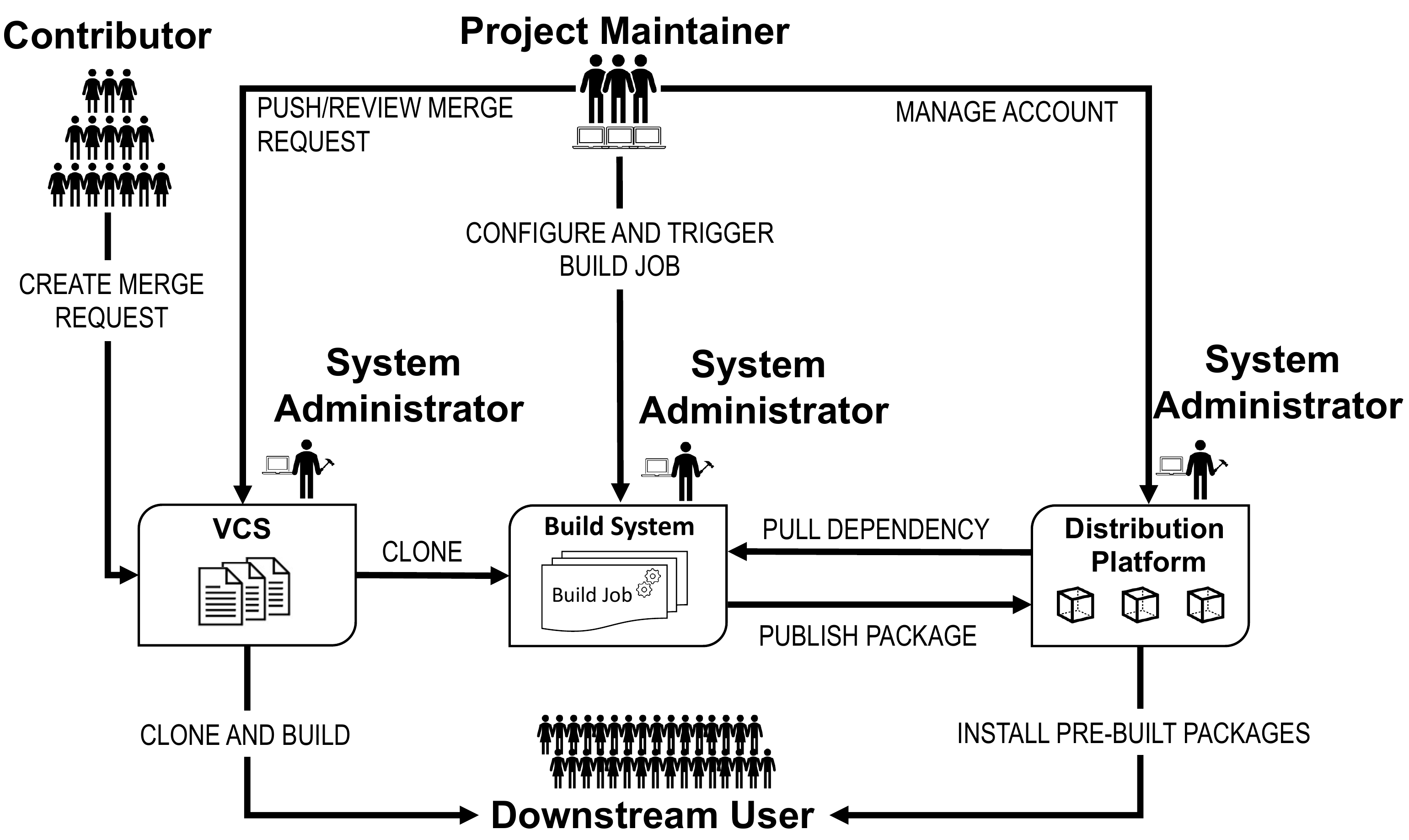}
    \caption{Stakeholders, systems and dataflows  related to
    the development, build and distribution of \ac{OSS} artifacts~\cite{ladisa2022taxonomy}.}
    \label{fig:sdlc}
\end{figure}

The \ac{OSS} supply chain denotes all the systems and stakeholders involved
in the development, build, and distribution of \ac{OSS} artifacts to downstream users. 
Figure~\ref{fig:sdlc} describes at high-level the common \ac{OSS} development model.


The systems considered comprise \ac{VCS}, build systems, and distribution platforms (e.g., package
repositories). 

\textbf{\acl{VCS}s}
host the source code of the \ac{OSS} project, as well as
metadata, configuration files, and other resources. They track and manage all
the changes of the codebase throughout the development process.
Plain \ac{VCS}s like Git do not require user authentication, but
complementary tools (e.g., GitHub) offer additional functionalities
(e.g., issue trackers) or security controls (e.g., authentication, review workflows).

\textbf{Build Systems}
consume the project's code to produce a binary artifact, e.g., an
executable or compressed archive, which can be distributed to downstream users. The build commonly involves so-called dependency or
package managers (e.g., pip for Python) which determine and download all dependencies necessary for the build to
succeed.
\ac{CI}/\ac{CD} pipelines
are often used to automate the build, test, and deployment of project artifacts.

\textbf{Distribution Platforms}
distribute pre-built \ac{OSS} artifacts to downstream users, e.g.,
upon the execution of package managers or through manual download. They include not only the well-known public package repositories like PyPI
or Maven Central but also internal and external mirrors, \ac{CDN} or proxies.

\textbf{Workstations of \ac{OSS} Maintainers and Administrators.}
\ac{OSS} project maintainers and administrators of the abovementioned systems
have privileged access to sensitive resources, e.g., the codebase, a build
system's web interface, or a package repository's database. Therefore, their
workstations are in the scope of the attack scenario.


Concerning the stakeholders involved in the \ac{OSS} supply chain, they include contributors, project maintainers, system and service administrators, and downstream users.


\textbf{Contributors}
contribute to an \ac{OSS} project with limited 
access to project resources. 
Common way of contributions to the \ac{VCS} involve the submission of merge requests. 

\textbf{Project Maintainers}
\label{sec:bg:maintainers}
have privileged access to project resources. For example, they are in charge of reviewing and integrating
contributors' merge requests, configure build systems and trigger build jobs, or
deploy ready-made artifacts on package repositories.

\textbf{System and Service Administrators}
\label{sec:bg:admin}
have the responsibility to configure, maintain, and operate any of the
above-mentioned systems. These stakeholders can include, e.g., employees of 3rd-party \ac{VCS} hosting
providers, members of \ac{OSS} foundations that operate private build systems for
their projects, or employees of companies running package repositories (like npm or private mirrors).

\textbf{Downstream Users}
consume \ac{OSS} project artifacts. They can opt to consume the source code from \ac{VCS} (and then build it themselves) or, as is more common, download pre-built versions from distribution platforms.  
In the context of downstream development projects, the download is
typically automated by package managers like pip or npm, which help in automatically 
identifying and obtaining the direct and transitive dependencies for a certain project. 

Both systems and stakeholders have to be considered as roles,
multiple of which can be exercised by a single host, individual, or third party service. For example, maintainers of an
 \ac{OSS} project typically consume artifacts of other projects

\subsection{Risks of Open-Source Software Supply Chains}


The above-described systems are inherently distributed, and the stakeholders are
partly unknown or anonymous. The real identities of project collaborators, both contributors and maintainers, are
not necessarily known. Accounts, including anonymous ones, gain trust through
continued contributions of quality (meritocracy).

Each open-source component used comes with its own systems and stakeholders, thereby multiplying an attack surface having both technical and social
facets. As in other adversarial contexts, attackers require
finding single weaknesses, while defenders need to cover the whole attack
surface, which in this case spans the whole supply chain.

Even heavily used open-source projects receive only little
funding and contributions, making it difficult 
for maintainers to securely run projects or promptly react to security incidents.
Moreover, the larger the user base (direct and indirect) the more attractive an open-source project
becomes for attackers. 

Downstream consumers have no control over and limited visibility into given
projects' security practices. The sheer number of 
dependencies 
makes rigorous reviews impractical for a
given consumer, forcing them to trust the community for a timely detection of
vulnerabilities and attacks.

\subsection{Attacker Model}
\label{section:attackermodel}

To identify attack vectors related to \ac{OSS} supply chains, we make the following assumptions on the attacker.



Primary goal of the attacker is to \textbf{insert} malicious code in
open-source artifacts such that it is executed in the context of downstream
projects, e.g., during its development or runtime.
Targeted assets can
belong both to developers of downstream software projects, or their end-users,
depending on the attacker's specific intention. In fact, the focus of the
taxonomy is not on \textit{what} malicious code does, but \textit{how} attackers
place it in upstream projects.

\begin{quote}
  The focus of the
taxonomy is not \textit{what} malicious code does, but \textit{how} attackers
place it in upstream projects.
\end{quote}


Initially, attackers only have access to publicly available information and publicly
accessible resources, which they can collect and analyze following the
\ac{OSINT} 
approach. Of course, due to the nature of
open-source projects, many project details are freely accessible, e.g., project
dependencies, build information, or commit and merge request histories.
Attackers can interact with any of the stakeholders and resources depicted in
Figure~\ref{fig:sdlc}, e.g., to communicate with maintainers using merge requests
or issue trackers or to create fake accounts and projects.

Attackers can target any kind of project (e.g., libraries, word processors).
    Downstream consumers can be affected directly or indirectly and in addition the scope of the attacker 
    can be either to reach a large pool of consumers or to target a specific group. This is possible by conditioning the execution of malicious code depending, e.g., on
    the lifecycle phase (install, test, etc.), application state, operating system,
    or properties of the downstream component it has been integrated
    into~\cite{ohm2020backstabbers}.

\section{Our start: Taxonomy of Attacks}\label{sec:taxonomy}

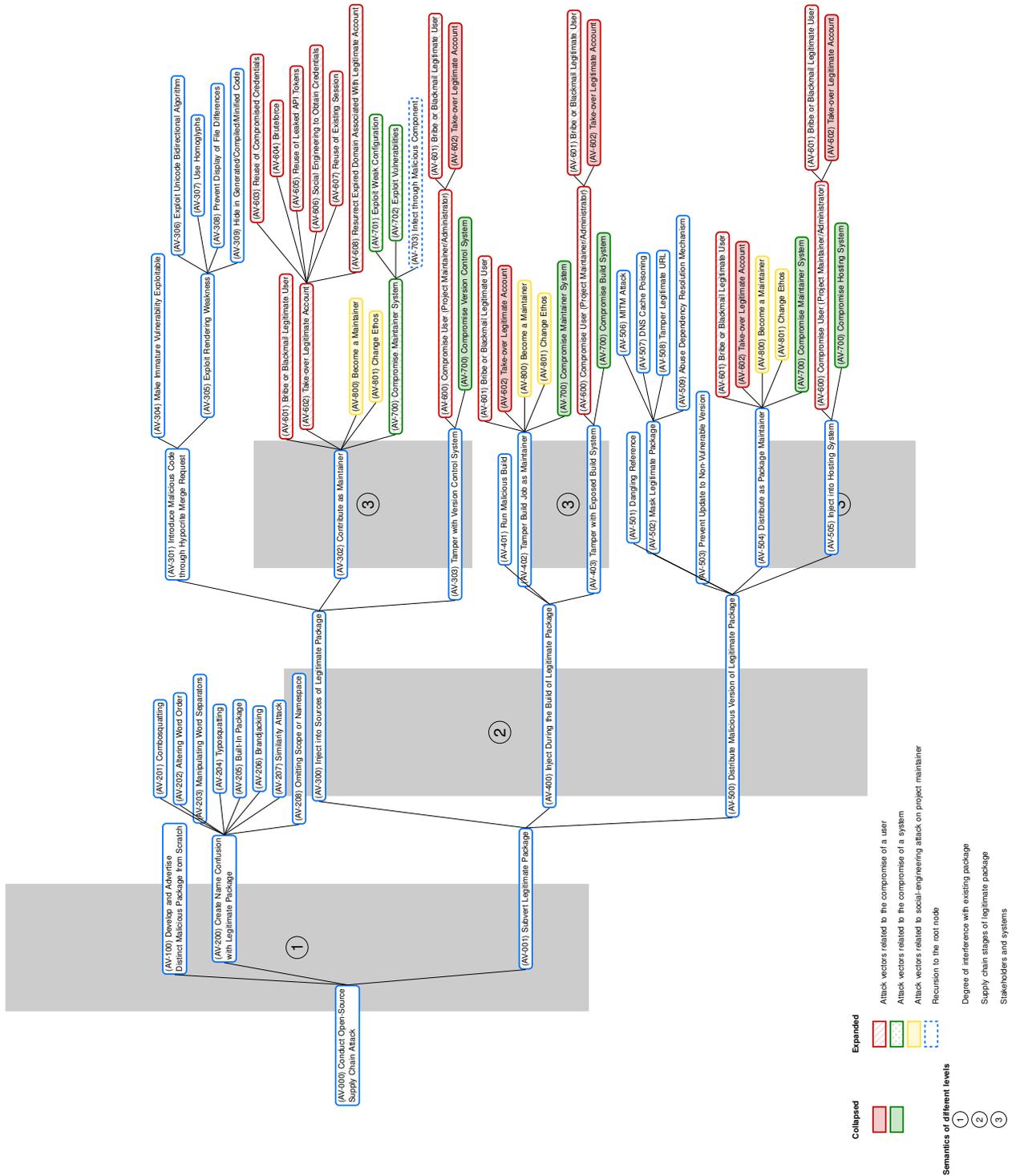
\begin{figure*}[htp]

	\begin{adjustbox}{scale=0.42,angle=90}
	
	\sbox0{%
	\begin{forest}
		/tikz/every node/.append style={font=\normalsize},
		for tree={%
		  treenode,
		parent anchor=east,child anchor=west,
		  grow'=east}
		[ (AV-000) Conduct Open-Source \\ Supply Chain Attack 
	[ (AV-100) Develop and Advertise \\Distinct Malicious Package from Scratch]
	[ (AV-200) Create Name Confusion \\with Legitimate Package 
	[ (AV-201) Combosquatting ]
	[ (AV-202) Altering Word Order ]
	[ (AV-203) Manipulating Word Separators ]
	[ (AV-204) Typosquatting ]
	[ (AV-205) Built-In Package ] 
	[ (AV-206) Brandjacking ]
	[ (AV-207) Similarity Attack ]
	[ (AV-208) Omitting Scope or Namespace ]
	]
	[ (AV-001) Subvert Legitimate Package
	[ (AV-300) Inject into Sources of Legitimate Package 
	[ (AV-301) Introduce Malicious Code \\ through Hypocrite Merge Request 
	[ (AV-304) Make Immature Vulnerability Exploitable ]
	[ (AV-305) Exploit Rendering Weakness  
	[ (AV-306) Exploit Unicode Bidirectional Algorithm ]
	[ (AV-307) Use Homoglyphs ]
	[ (AV-308) Prevent Display of File Differences ]
	[ (AV-309) Hide in Generated/Compiled/Minified Code ]
	]
	]
	[ (AV-302) Contribute as Maintainer 
	[ (AV-601) Bribe or Blackmail Legitimate User ,for tree={draw=bostonuniversityred}, for tree={draw=bostonuniversityred, pattern=north east lines, pattern color=bostonuniversityred!20},]
	[ (AV-602) Take-over Legitimate Account,for tree={draw=bostonuniversityred, pattern=north east lines, pattern color=bostonuniversityred!20},
	[ (AV-603) Reuse of Compromised Credentials]
	[ (AV-604) Bruteforce ]
	[ (AV-605) Reuse of Leaked API Tokens]
	[ (AV-606) Social Engineering to Obtain Credentials]
	[ (AV-607) Reuse of Existing Session ]
	[ (AV-608) Resurrect Expired Domain Associated With Legitimate Account ]
	]
	[ (AV-800) Become a Maintainer, top color=bananayellow!20,bottom color=bananayellow!20,draw=bananayellow]
	[ (AV-801) Change Ethos, top color=bananayellow!20,bottom color=bananayellow!20,draw=bananayellow]
	[ (AV-700) Compromise Maintainer System , for tree={draw=ao(english), pattern=crosshatch dots, pattern color=ao(english)!20}, 
	[ (AV-701) Exploit Weak Configuration]
	[ (AV-702) Exploit Vulnerabilities ]
	[ (AV-703) Infect through Malicious Component , draw=brandeisblue, dashed, fill=white]
	]
	]
	[ (AV-303) Tamper with Version Control System
	[ (AV-600) Compromise User (Project Maintainer/Administrator), for tree={draw=bostonuniversityred, pattern=north east lines, pattern color=bostonuniversityred!20},
	[ (AV-601) Bribe or Blackmail Legitimate User ]
	[ (AV-602) Take-over Legitimate Account , for tree={draw=bostonuniversityred, top color=bostonuniversityred!20,, bottom color=bostonuniversityred!20}
	]
	]
	[ (AV-700) Compromise Version Control System ,for tree={draw=ao(english), top color=ao(english)!20, bottom color=ao(english)!20},
	]
	]
	]
	[ (AV-400) Inject During the Build of Legitimate Package 
	[ (AV-401) Run Malicious Build ]
	[ (AV-402) Tamper Build Job as Maintainer 
	[ (AV-601) Bribe or Blackmail Legitimate User ,for tree={draw=bostonuniversityred, pattern=north east lines, pattern color=bostonuniversityred!20},]
	[ (AV-602) Take-over Legitimate Account ,for tree={draw=bostonuniversityred, top color=bostonuniversityred!20,, bottom color=bostonuniversityred!20},
	]
	[ (AV-800) Become a Maintainer, top color=bananayellow!20,bottom color=bananayellow!20,draw=bananayellow]
	[ (AV-801) Change Ethos, top color=bananayellow!20,bottom color=bananayellow!20,draw=bananayellow]
	[ (AV-700) Compromise Maintainer System , for tree={draw=ao(english), top color=ao(english)!20, bottom color=ao(english)!20},
	]
	]
	[ (AV-403) Tamper with Exposed Build System 
	[ (AV-600) Compromise User (Project Maintainer/Administrator),for tree={draw=bostonuniversityred, pattern=north east lines, pattern color=bostonuniversityred!20},
	[ (AV-601) Bribe or Blackmail Legitimate User ]
	[ (AV-602) Take-over Legitimate Account, for tree={draw=bostonuniversityred, top color=bostonuniversityred!20,, bottom color=bostonuniversityred!20},
	]
	]
	[ (AV-700) Compromise Build System, for tree={draw=ao(english), top color=ao(english)!20, bottom color=ao(english)!20},
	]
	]
	]
	[ (AV-500) Distribute Malicious Version of Legitimate Package 
	[ (AV-501) Dangling Reference ]
	[ (AV-502) Mask Legitimate Package 
	[ (AV-506) MITM Attack ]
	[ (AV-507) DNS Cache Poisoning ]
	[ (AV-508) Tamper Legitimate URL ]
	[ (AV-509) Abuse Dependency Resolution Mechanism]
	]
	[ (AV-503) Prevent Update to Non-Vulnerable Version ]
	[ (AV-504) Distribute as Package Maintainer 
	[ (AV-601) Bribe or Blackmail Legitimate User ,for tree={draw=bostonuniversityred, pattern=north east lines, pattern color=bostonuniversityred!20}]
	[ (AV-602) Take-over Legitimate Account ,for tree={draw=bostonuniversityred, top color=bostonuniversityred!20,, bottom color=bostonuniversityred!20},
	]
	[ (AV-800) Become a Maintainer, top color=bananayellow!20,bottom color=bananayellow!20,draw=bananayellow]
	[ (AV-801) Change Ethos, top color=bananayellow!20,bottom color=bananayellow!20,draw=bananayellow]
	[ (AV-700) Compromise Maintainer System , for tree={draw=ao(english), top color=ao(english)!20, bottom color=ao(english)!20},
	]
	]
	[ (AV-505) Inject into Hosting System 
	[ (AV-600) Compromise User (Project Maintainer/Administrator),for tree={draw=bostonuniversityred, pattern=north east lines, pattern color=bostonuniversityred!20},
	[ (AV-601) Bribe or Blackmail Legitimate User ]
	[ (AV-602) Take-over Legitimate Account , for tree={draw=bostonuniversityred, top color=bostonuniversityred!20,, bottom color=bostonuniversityred!20}
	]
	]
	[ (AV-700) Compromise Hosting System, for tree={draw=ao(english), top color=ao(english)!20, bottom color=ao(english)!20},
	]
	]
	]
	]
	]
		\end{forest}}%

	\sbox1{
	  
	\begin{tabular}{ccl}
		
		\textbf{Collapsed} & \textbf{Expanded} & \\
		\\
		\tikz{\draw[bostonuniversityred,ultra thick, fill=bostonuniversityred!20] (0,0) rectangle (1,0.5)} &\tikz{\draw[bostonuniversityred,pattern=north east lines, pattern color=bostonuniversityred!20,ultra thick] (0,0) rectangle (1,0.5)} & Attack vectors related to the compromise of a user\\
		\tikz{\draw[ao(english),fill=ao(english)!20,ultra thick] (0,0) rectangle (1,0.5)} &\tikz{\draw[ao(english),pattern=crosshatch dots, pattern color=ao(english)!20,ultra thick] (0,0) rectangle (1,0.5)} & Attack vectors related to the compromise of a system\\
		 &\tikz{\draw[bananayellow,fill=bananayellow!20,ultra thick] (0,0) rectangle (1,0.5)} & Attack vectors related to social-engineering attack on project maintainer\\
		 &\tikz{\draw[brandeisblue,dashed,ultra thick] (0,0) rectangle (1,0.5)} & Recursion to the root node\\
		 \textbf{Semantics of different levels} & &\\
		 \tikz{\node[draw,circle] {1};} & & Degree of interference with existing package
		\\
		\tikz{\node[draw,circle] {2};} & & Supply chain stages of legitimate package
		\\
		\tikz{\node[draw,circle] {3};} & & Stakeholders and systems
		\\
		
		\end{tabular}

		}%
		\sbox2{
			
			\tikz{\filldraw[black!20] (0,10) rectangle (5,33)}
	
		}
	
		\sbox3{
			\tikz{\filldraw[black!20] (0,0) rectangle (5,23)}
		}
		\sbox4{
		  
			\tikz{\filldraw[black!20] (0,0) rectangle (5,3.2)}
		}
	
		\sbox5{
			\tikz{\filldraw[black!20] (0,0) rectangle (5,6)}
		}
		\sbox6{
			
			\tikz{\filldraw[black!20] (0,0) rectangle (5,8.6)}
		}
	
		\sbox7{
			\tikz{\node[draw,circle] {\Huge{3}};}
		}
		\sbox8{
			\tikz{\node[draw,circle] {\Huge{2}};}
		}
		\sbox9{
			\tikz{\node[draw,circle] {\Huge{1}};}
		}

	\begin{tikzpicture}
	  \node (forest){\usebox0};
	  \node[] at (-16,-17) {\usebox1};
	  \begin{scope}[on background layer]
		\node[rounded corners] at (-15.5, 8) {\usebox2 } ;
		\node[rounded corners] at (-15.5, 8) {\usebox9 } ;
		\node[rounded corners] at (-7, -3) {\usebox3 } ;
		\node[rounded corners] at (-7, 0) {\usebox8 } ;
		\node[rounded corners] at (2, -2.7) {\usebox4 } ;
		\node[rounded corners] at (2, -2.7) {\usebox7 } ;
		\node[rounded corners] at (2, -13.4) {\usebox5 } ;
		\node[rounded corners] at (2, -13.4) {\usebox7 } ;
		\node[rounded corners] at (2, 5.4) {\usebox6 } ;
		\node[rounded corners] at (2, 5.2) {\usebox7 } ;
	\end{scope}
	  
	\end{tikzpicture}

	\end{adjustbox}

		\caption{Refined version of the taxonomy for \ac{OSS} supply chain attacks~\cite{ladisa2022taxonomy}.
		 }
		\label{fig:attacktreetaxonomy}
	
	\end{figure*}


    This section present the taxonomy consisting of 117 unique attack vectors, collected through the review of scientific and grey literature (details in~\cite{ladisa2022taxonomy}). 
    Such taxonomy takes the form of an \textit{attack tree} to systematically represent the attacker goals and techniques. In an attack tree, the root node represents the attacker's top-level goal, which is iteratively refined by its children into subgoals.

    To create the taxonomy depicted in Figure~\ref{fig:attacktreetaxonomy} we conducted a \ac{SLR} of both scientific and grey literature. We have done our best to collect as many resources as possible, although we refrain from claiming to have achieved completeness. 
    By conducting a user survey, the taxonomy has been validated and assessed by 17 experts in software supply chain security and 134 software developers.

    
    \paragraph{Conducting an Open-Source Supply Chain Attack} 
    This is the attackers' top-level goal and happens by injecting
    malicious code
    into an \ac{OSS} project such that it is downloaded by
    downstream consumers, and executed upon installation or at runtime.

    The entire taxonomy unfolding below this high-level goal is depicted in
    Figure~\ref{fig:attacktreetaxonomy} and summarized in the following.
    The 1st-level child nodes of the tree reflect different degrees of interference
    with existing packages.
    
    


    \paragraph{Develop and Advertise Distinct Malicious Package from Scratch}

    This node covers the creation of a brand new \ac{OSS} project, with the intention
    to use it for spreading malicious code from the beginning or at a later point
    in time. Besides creating the project, the attacker is required to advertise the
    project to attract victims. Real-world examples affect PyPI, npm, Docker Hub or
    NuGet.
    
    
    
    
    \paragraph{Create Name Confusion with Legitimate Package}

    This node covers attacks that consist of creating project or artifact names that resemble
    legitimate ones, suggest trustworthy authors, or play with common naming
    patterns. Once a suitable name is found, the malicious artifact is deployed,
    e.g., in a source or package repository, in the hope of being consumed by 
    downstream users. As the deployment does not interfere with the resources of the
    project that inspired the name (e.g., legitimate code repository, maintainer
    accounts) the attack is relatively cheap.
    
    Child nodes of this attack vector relate to sub-techniques applying different
    modifications to the legitimate project name:
    \emph{Combosquatting} 
     adds pre or
    post-fixes, e.g., to indicate project maturity (\texttt{dev} or
    \texttt{rc}) or platform compatibility (\texttt{i386}).
    \emph{Altering Word Order} 
     re-arranges
    the word order (\texttt{test-vision-client} vs.
    \texttt{client-vision-test}).
    \emph{Manipulating Word Separators} 
     alters or adds word separators like hyphens (\texttt{setup-tools} vs.
    \texttt{setuptools}).
    \emph{Typosquatting} 
    exploits typographical errors (\texttt{dajngo} vs. \texttt{django}).
    \emph{Built-In Package} 
    replicates well-known names from other contexts, e.g., built-in packages or modules of a
    programming language (\texttt{subprocess} for Python).
    %
    \emph{Brandjacking} 
    includes the name of popular brands/organizations (e.g., \texttt{twilio, aws})
    to suggest that such package comes from a trustworthy author (\texttt{twilio-npm}).
    \emph{Similarity Attack} 
    creates a misleading name in a way different from the
    previous categories (\texttt{request} vs. \texttt{requests}).
    
    \paragraph{Subvert Legitimate Package} 
    This node covers all attacks aiming to corrupt an
    existing, legitimate project, which requires compromising one or more of its
    numerous resources depicted in Figure~\ref{fig:sdlc}. As a result, this subtree
    is much larger compared to the previous ones, especially because subtrees related to
    user and system compromises occur multiple times in the different supply chain
    stages. 
    
    The remainder of this section is dedicated to sub-techniques of this
    first-level node.
    
    \textit{Inject into Sources of Legitimate Package}
    \label{injectintosources}
    relates to the injection of malicious code into a project's
    codebase. From the attacker's point of view, this has the advantage to affect all downstream
    users, no matter whether they consume sources or pre-built binary artifacts (as
    part of the codebase, the malicious code will be included during project builds
    and binary artifact distribution). 
    This vector has several sub-techniques. Taking the role of contributors,
    attackers can use \emph{hypocrite merge requests} to turn immature
    vulnerabilities into exploitable ones, 
    or exploit IDE rendering weaknesses to hide malicious code, e.g., through the use of Unicode
    homoglyphs and control characters 
    , or the hiding and suppression of code differences.
    %
    To \emph{contribute as maintainer} requires to obtain the privileges necessary
    for altering the legitimate project's codebase, which can be achieved in
    different ways: using \ac{SE} techniques on legitimate project maintainers,
    \emph{changing the ethos} (e.g., as in the case of protestwares), 
    \emph{taking over legitimate accounts} (e.g., reusing compromised
    credentials
    , or \emph{compromising the maintainer
    system} (e.g., exploiting vulnerabilities). The 
    latter can also be achieved through a malicious (\ac{OSS}) component, e.g., IDE
    plugin, which is reflected through a recursive reference to the root node.
    The legitimate project's codebase can also be altered by \emph{tampering with
    its \ac{VCS}}, thus, bypassing a project's established contribution workflows. For
    instance, by compromising system user
    accounts
    , or by exploiting
    configuration/software
    vulnerabilities
    , an attacker could access the codebase in insecure ways.

    \textit{Inject During the Build of Legitimate Package}
    \label{injectduringbuild}
    Greatly facilitated by language-specific package managers like Maven or Gradle
    for Java, it became common to download pre-built components from package
    repositories rather than \ac{OSS} project's source code from its \ac{VCS}.
    Therefore, the injection of malicious code can happen during the build of such
    components before their
    publication.
    Though the spread is limited compared to injecting into sources, the advantage
    for the attacker is that the detection of malicious code inside pre-built
    packages is typically more difficult, especially for compiled programming
    languages.
    One sub-technique is \emph{running a malicious build job} to tamper with system
    resources shared between build jobs of multiple
    projects
    (e.g., the infection of Java archives in NetBeans projects). 
    An attacker can also \emph{tamper the build job as maintainer}, e.g., by taking
    over legitimate maintainer accounts, becoming a maintainer, or compromising their
    systems (e.g., XCodeGhost malware). 
    Similarly, the
    attacker could compromise build systems, especially online accessible ones, e.g., by
    compromising administrator accounts 
    or exploiting vulnerabilities.
    
    \textit{Distribute Malicious Version of Legitimate Package}
    Pre-built components are often hosted on well-known package repositories like
    PyPI or npm, but also on less popular repositories with a narrower scope. In
    addition, the components can be mirrored remotely or locally, made available
    through \ac{CDN}s (e.g., in the case of JavaScript libraries), or cached in
    proxies. This attack vector and its sub-techniques cover all cases where
    attackers tamper with mechanisms and systems involved in the hosting,
    distribution, and download of pre-built packages.
    \emph{Dangling references} (re)use resource identifiers of orphaned
    projects, 
    e.g., names or URLs.
    \emph{Mask legitimate package} 
    targets package
    name or URL resolution mechanisms and download connections. Their goal is the
    download of malicious packages by compromising resources external to the
    legitimate project. This includes \ac{MITM} attacks, DNS cache poisoning, or
    tampering with legitimate URLs directly at the
    client. 
    Particularly, package managers follow a
    (configurable) resolution strategy to decide which package version to download,
    from where, and the order of precedence when contacting multiple repositories.
    Attackers can \emph{abuse such resolution mechanisms} and their
    configurations.
    Attackers can also \emph{prevent updates to non-vulnerable versions} by
    manipulating package metadata
    , e.g., by indicating
    an unsatisfiable dependency for newer versions of a legitimate package.
    Finally, the involvement of systems and users in package distribution results in
    attack vectors similar to previous ones. Attackers can take the role of
    legitimate maintainers, thus, \textit{distribute as maintainer}, e.g., by taking
    over package maintainer accounts (e.g., \texttt{eslint}), the
    second most common attack vector after typosquatting. 
    They can also compromise maintainer systems, or directly \emph{inject into the
    hosting system}, e.g., by compromising administrator accounts or 
    exploiting vulnerabilities. 

    \subsection{Remark}
    While the taxonomy presented is largely agnostic, some attack vectors are specific to certain ecosystems.
   \textit{Abuse Dependency Resolution
    Mechanism} attacks depend on the approach and strategy used by the respective
    package manager to resolve and download declared dependencies from internal and
    external repositories. For instance, Maven, npm, pip, NuGet or Composer were
    affected by the dependency confusion attack, while Go and Cargo were
    not~\cite{dependencyconfusionschibsted}. Several attacks below \textit{Exploit
    Rendering Weakness} depend on the interpretation and visualization of (Unicode)
    characters by user interfaces and
    compiler/interpreters~\cite{boucher2021trojan}. Also name confusion attacks
    need to consider ecosystem specificities, especially \textit{Built-In Packages}.

    \definecolor{syellowcolor}{rgb}{0.98,0.64,0.098}

    \begin{tcolorbox}[enhanced jigsaw,breakable,pad at break*=1mm,colback=syellowcolor!5!white,colframe=syellowcolor,title=\textbf{Pills of History of Supply Chain Attacks}]
      Since its first appearance in 1984 with the publication of Ken Thompson's \textit{Reflections on Trusting Trusts}, the discussion around software supply chain security has kept growing, with a larger increase in recent years.
In terms of attacks, to the best of our knowledge, the first known \ac{OSS} supply chain attacks date back to 2010 (cf. Figure~\ref{fig:refs-years-attacks}). The first consists of the implantation of a backdoor in the source code of ProFTPD1.3.3c~\cite{zdnetOpensourceProFTPD} while the second is the well-known case of Operation Aurora~\cite{mcclure2010protecting}.
Until 2016, attacks mainly consist in the injection of malicious code into source code or during builds, or the deployment of malicious versions of legitimate softwares in hosting systems. It's in 2017 that we detect the first attack that exploited the attack vector \textit{Create Name Confusion with Legitimate Package}, which probably is the most exploited attack vector today. In particular, this first attack involved uploading packages to PyPI with a name squatting that of popular projects such as urllib3 and acquisition (using different sub-techniques). Prior to this attack, there's just one non-peer-reviewed work discussing the attack vector \textit{Create Name Confusion with Legitimate Package}, that is Nikolai Tschacher's Master thesis~\cite{tschacher2016typosquatting} published in 2016, one year before the first attack of this kind.
    \end{tcolorbox}

\section{A rapid recovery: Safeguards}

\begin{table*}[!hbtp]

    \centering
    \resizebox{\textwidth}{!}{%
    \begin{tabular}{r|llllllllll|}
    \toprule
    
    \multicolumn{2}{l}{} & \multicolumn{4}{c}{\textbf{Control Type}} & \multicolumn{3}{c}{\textbf{Stakeholders Involved}} & \multicolumn{1}{l}{}\\ \cmidrule(lr){3-6} \cmidrule(lr){7-9}
    
    \multicolumn{1}{c}{\multirow{-2}{*}{\textbf{Safeguard}}}& \multicolumn{1}{l}{\rotatebox[origin=c]{70}{\textbf{Utility-to-Cost}}}  & \multicolumn{1}{l}{\rotatebox[origin=c]{70}{\textbf{Directive}}} & \multicolumn{1}{l}{\rotatebox[origin=c]{70}{\textbf{Preventive}}} & \multicolumn{1}{l}{\rotatebox[origin=c]{70}{\textbf{Detective}}} & \multicolumn{1}{l}{\rotatebox[origin=c]{70}{\textbf{Corrective}}} & \multicolumn{1}{c}{\rotatebox[origin=c]{70}{\textbf{\ac{OSS} Maintainer}}} & \multicolumn{1}{c}{\rotatebox[origin=c]{70}{\textbf{3P Service Prov.}}} & \multicolumn{1}{c}{\rotatebox[origin=c]{70}{\textbf{\ac{OSS} Consumer}}}& \multicolumn{1}{c}{\multirow{-2}{*}{\textbf{Attack-Vector Addressed}}} \\ 
    
    \midrule
    \multicolumn{1}{p{5.4cm}}{Protect production branch}& \multicolumn{1}{c}{2.10} & \multicolumn{1}{c}{} & \multicolumn{1}{c}{\checkmark} & \multicolumn{1}{c}{\checkmark} & \multicolumn{1}{c}{}& \multicolumn{1}{c}{$\bullet$}& \multicolumn{1}{c}{}& \multicolumn{1}{c}{} &  \multicolumn{1}{c}{AV-301, AV-302} \\
    \multicolumn{1}{p{5.4cm}}{Remove un-used dependencies} & \multicolumn{1}{c}{2.05}  & \multicolumn{1}{c}{} & \multicolumn{1}{c}{\checkmark} & \multicolumn{1}{c}{} & \multicolumn{1}{c}{} & \multicolumn{1}{c}{}& \multicolumn{1}{c}{}& \multicolumn{1}{c}{$\bullet$}&  \multicolumn{1}{c}{AV-001}\\
    \multicolumn{1}{p{5.4cm}}{Version pinning}& \multicolumn{1}{c}{1.68} & \multicolumn{1}{c}{} & \multicolumn{1}{c}{\checkmark} & \multicolumn{1}{c}{} & \multicolumn{1}{c}{} & \multicolumn{1}{c}{}& \multicolumn{1}{c}{}& \multicolumn{1}{c}{$\bullet$}&  \multicolumn{1}{c}{AV-001}\\
    \multicolumn{1}{p{5.4cm}}{Dependency resolution rules}& \multicolumn{1}{c}{1.58} & \multicolumn{1}{c}{} & \multicolumn{1}{c}{\checkmark} & \multicolumn{1}{c}{} & \multicolumn{1}{c}{}& \multicolumn{1}{c}{}& \multicolumn{1}{c}{}& \multicolumn{1}{c}{$\bullet$} &  \multicolumn{1}{c}{AV-501, AV-508, AV-509}\\
    \multicolumn{1}{p{5.4cm}}{User account management}& \multicolumn{1}{c}{1.50} & \multicolumn{1}{c}{} & \multicolumn{1}{c}{\checkmark} & \multicolumn{1}{c}{} & \multicolumn{1}{c}{\checkmark}& \multicolumn{1}{c}{$\bullet$}& \multicolumn{1}{c}{$\bullet$}& \multicolumn{1}{c}{}  &  \multicolumn{1}{c}{AV-302,AV-402,AV-504,AV-600}\\
    \multicolumn{1}{p{5.4cm}}{Secure authentication (e.g., \ac{MFA}, password recycle, session timeout, token protection)}& \multicolumn{1}{c}{1.48} & \multicolumn{1}{c}{} & \multicolumn{1}{c}{\checkmark} & \multicolumn{1}{c}{} & \multicolumn{1}{c}{}& \multicolumn{1}{c}{$\bullet$}& \multicolumn{1}{c}{$\bullet$}& \multicolumn{1}{c}{} &  \multicolumn{1}{c}{AV-*00 $\rightarrow$ AV-602}\\
    \multicolumn{1}{p{5.4cm}}{Use of security, quality and health metrics} & \multicolumn{1}{c}{1.35} & \multicolumn{1}{l}{\checkmark} & \multicolumn{1}{l}{\checkmark} & \multicolumn{1}{l}{} & \multicolumn{1}{l}{}& \multicolumn{1}{c}{$\bullet$}& \multicolumn{1}{c}{$\bullet$}& \multicolumn{1}{c}{$\bullet$} &  \multicolumn{1}{c}{AV-000}\\
    \multicolumn{1}{p{5.4cm}}{Typo guard/Typo detection}& \multicolumn{1}{c}{1.34} & \multicolumn{1}{c}{} & \multicolumn{1}{c}{\checkmark} & \multicolumn{1}{c}{\checkmark} & \multicolumn{1}{c}{}& \multicolumn{1}{c}{}& \multicolumn{1}{c}{$\bullet$}& \multicolumn{1}{c}{$\bullet$} &  \multicolumn{1}{c}{AV-200}\\
    \multicolumn{1}{p{5.4cm}}{Use minimal set of trusted build dependencies in the release job}& \multicolumn{1}{c}{1.32} & \multicolumn{1}{c}{} & \multicolumn{1}{c}{\checkmark} & \multicolumn{1}{c}{} & \multicolumn{1}{c}{}& \multicolumn{1}{c}{$\bullet$}& \multicolumn{1}{c}{}& \multicolumn{1}{c}{} &  \multicolumn{1}{c}{AV-400} \\
    \multicolumn{1}{p{5.4cm}}{Integrity check of dependencies through cryptographic hashes}& \multicolumn{1}{c}{1.32} & \multicolumn{1}{c}{} & \multicolumn{1}{c}{} & \multicolumn{1}{c}{\checkmark} & \multicolumn{1}{c}{}& \multicolumn{1}{c}{}& \multicolumn{1}{c}{}& \multicolumn{1}{c}{$\bullet$} &  \multicolumn{1}{c}{AV-400, AV-500}\\ 
    \multicolumn{1}{p{5.4cm}}{Maintain detailed \ac{SBOM} and perform \ac{SCA}} & \multicolumn{1}{c}{1.24} & \multicolumn{1}{l}{} & \multicolumn{1}{c}{\checkmark} & \multicolumn{1}{c}{\checkmark} & \multicolumn{1}{l}{} & \multicolumn{1}{c}{$\bullet$}& \multicolumn{1}{c}{$\bullet$}& \multicolumn{1}{c}{$\bullet$} &  \multicolumn{1}{c}{AV-000}\\ 
    \multicolumn{1}{p{5.4cm}}{Ephemeral build environment} & \multicolumn{1}{c}{1.24} & \multicolumn{1}{c}{} & \multicolumn{1}{c}{\checkmark} & \multicolumn{1}{c}{} & \multicolumn{1}{c}{}& \multicolumn{1}{c}{$\bullet$}& \multicolumn{1}{c}{}& \multicolumn{1}{c}{} &  \multicolumn{1}{c}{AV-400} \\ 
    \multicolumn{1}{p{5.4cm}}{Prevent script execution}& \multicolumn{1}{c}{1.23} & \multicolumn{1}{c}{} & \multicolumn{1}{c}{\checkmark} & \multicolumn{1}{c}{} & \multicolumn{1}{c}{} & \multicolumn{1}{c}{}& \multicolumn{1}{c}{}& \multicolumn{1}{c}{$\bullet$}&  \multicolumn{1}{c}{AV-000}\\
    \multicolumn{1}{p{5.4cm}}{Pull/Merge request review}& \multicolumn{1}{c}{1.21} & \multicolumn{1}{c}{} & \multicolumn{1}{c}{\checkmark} & \multicolumn{1}{c}{} & \multicolumn{1}{c}{} & \multicolumn{1}{c}{$\bullet$}& \multicolumn{1}{c}{}& \multicolumn{1}{c}{}&  \multicolumn{1}{c}{AV-301, AV-302} \\
    \multicolumn{1}{p{5.4cm}}{Restrict access to system resources of code executed during each build steps}& \multicolumn{1}{c}{1.21} & \multicolumn{1}{c}{} & \multicolumn{1}{c}{\checkmark} & \multicolumn{1}{c}{} & \multicolumn{1}{c}{}& \multicolumn{1}{c}{$\bullet$}& \multicolumn{1}{c}{}& \multicolumn{1}{c}{} &  \multicolumn{1}{c}{AV-400} \\
    \multicolumn{1}{p{5.4cm}}{Code signing} & \multicolumn{1}{c}{1.19} & \multicolumn{1}{c}{} & \multicolumn{1}{c}{} & \multicolumn{1}{c}{\checkmark} & \multicolumn{1}{l}{}& \multicolumn{1}{c}{$\bullet$}& \multicolumn{1}{c}{$\bullet$}& \multicolumn{1}{c}{$\bullet$} &  \multicolumn{1}{c}{AV-200, AV-500} \\ 
    \multicolumn{1}{p{5.4cm}}{Integrate Open-Source vulnerability scanner into CI/CD pipeline}& \multicolumn{1}{c}{1.15} & \multicolumn{1}{c}{} & \multicolumn{1}{c}{} & \multicolumn{1}{c}{\checkmark} & \multicolumn{1}{c}{} & \multicolumn{1}{c}{}& \multicolumn{1}{c}{}& \multicolumn{1}{c}{$\bullet$}& \multicolumn{1}{c}{AV-000} \\
    \multicolumn{1}{p{5.4cm}}{Use of dedicated build service}& \multicolumn{1}{c}{1.09} & \multicolumn{1}{c}{} & \multicolumn{1}{c}{\checkmark} & \multicolumn{1}{c}{} & \multicolumn{1}{c}{}& \multicolumn{1}{c}{$\bullet$}& \multicolumn{1}{c}{}& \multicolumn{1}{c}{} &  \multicolumn{1}{c}{AV-400 $\rightarrow$ AV-700 } \\
    \multicolumn{1}{p{5.4cm}}{Preventive squatting the released packages}& \multicolumn{1}{c}{1.07} & \multicolumn{1}{c}{} & \multicolumn{1}{c}{\checkmark} & \multicolumn{1}{c}{} & \multicolumn{1}{c}{}& \multicolumn{1}{c}{$\bullet$}& \multicolumn{1}{c}{$\bullet$}& \multicolumn{1}{c}{}  &  \multicolumn{1}{c}{AV-200}\\ 
    \multicolumn{1}{p{5.4cm}}{Audit}& \multicolumn{1}{c}{1.05} & \multicolumn{1}{c}{\checkmark} & \multicolumn{1}{c}{} & \multicolumn{1}{c}{\checkmark} & \multicolumn{1}{c}{}& \multicolumn{1}{c}{$\bullet$}& \multicolumn{1}{c}{$\bullet$}& \multicolumn{1}{c}{}  &  \multicolumn{1}{c}{AV-000}\\ 
    \multicolumn{1}{p{5.4cm}}{Security assessment}& \multicolumn{1}{c}{1.05} & \multicolumn{1}{c}{} & \multicolumn{1}{c}{} & \multicolumn{1}{c}{\checkmark} & \multicolumn{1}{c}{}& \multicolumn{1}{c}{$\bullet$}& \multicolumn{1}{c}{$\bullet$}& \multicolumn{1}{c}{}  &  \multicolumn{1}{c}{AV-000}\\ 
    \multicolumn{1}{p{5.4cm}}{Vulnerability assessment}& \multicolumn{1}{c}{1.05} & \multicolumn{1}{c}{} & \multicolumn{1}{c}{} & \multicolumn{1}{c}{\checkmark} & \multicolumn{1}{c}{} & \multicolumn{1}{c}{$\bullet$}& \multicolumn{1}{c}{$\bullet$}& \multicolumn{1}{c}{} &  \multicolumn{1}{c}{AV-000}\\ 
    \multicolumn{1}{p{5.4cm}}{Penetration testing}& \multicolumn{1}{c}{1.05} & \multicolumn{1}{c}{} & \multicolumn{1}{c}{} & \multicolumn{1}{c}{\checkmark} & \multicolumn{1}{c}{}& \multicolumn{1}{c}{$\bullet$}& \multicolumn{1}{c}{$\bullet$}& \multicolumn{1}{c}{}  &  \multicolumn{1}{c}{AV-000}\\
    \multicolumn{1}{p{5.4cm}}{Reproducible builds}& \multicolumn{1}{c}{1.02} & \multicolumn{1}{l}{} & \multicolumn{1}{l}{} & \multicolumn{1}{c}{\checkmark} & \multicolumn{1}{l}{}& \multicolumn{1}{c}{$\bullet$}& \multicolumn{1}{c}{$\bullet$}& \multicolumn{1}{c}{$\bullet$} & \multicolumn{1}{c}{AV-400, AV-500}\\ 
    \multicolumn{1}{p{5.4cm}}{Isolation of build steps} & \multicolumn{1}{c}{1.00} & \multicolumn{1}{c}{} & \multicolumn{1}{c}{\checkmark} & \multicolumn{1}{c}{} & \multicolumn{1}{c}{}& \multicolumn{1}{c}{$\bullet$}& \multicolumn{1}{c}{}& \multicolumn{1}{c}{} &  \multicolumn{1}{c}{AV-400} \\
    \multicolumn{1}{p{5.4cm}}{Scoped packages}& \multicolumn{1}{c}{1.00} & \multicolumn{1}{c}{} & \multicolumn{1}{c}{\checkmark} & \multicolumn{1}{c}{} & \multicolumn{1}{c}{}& \multicolumn{1}{c}{$\bullet$}& \multicolumn{1}{c}{$\bullet$}& \multicolumn{1}{c}{}  &  \multicolumn{1}{c}{AV-509}\\
    \multicolumn{1}{p{5.4cm}}{Establish internal repository mirrors and reference one private feed, not multiple}& \multicolumn{1}{c}{0.97} & \multicolumn{1}{c}{} & \multicolumn{1}{c}{\checkmark} & \multicolumn{1}{c}{} & \multicolumn{1}{c}{} & \multicolumn{1}{c}{}& \multicolumn{1}{c}{}& \multicolumn{1}{c}{$\bullet$}&  \multicolumn{1}{c}{AV-501,AV-502, AV-504, AV-505}\\
    \multicolumn{1}{p{5.4cm}}{Application Security Testing } & \multicolumn{1}{c}{0.95}& \multicolumn{1}{c}{} & \multicolumn{1}{c}{} & \multicolumn{1}{c}{\checkmark} & \multicolumn{1}{c}{}& \multicolumn{1}{c}{}& \multicolumn{1}{c}{$\bullet$}& \multicolumn{1}{c}{$\bullet$} &  \multicolumn{1}{c}{AV-000}\\
    \multicolumn{1}{p{5.4cm}}{Establish vetting process for Open-Source components hosted in internal/public repositories}& \multicolumn{1}{c}{0.95} & \multicolumn{1}{c}{} & \multicolumn{1}{c}{\checkmark} & \multicolumn{1}{c}{} & \multicolumn{1}{c}{}& \multicolumn{1}{c}{}& \multicolumn{1}{c}{$\bullet$}& \multicolumn{1}{c}{$\bullet$} &  \multicolumn{1}{c}{AV-000}\\
    \multicolumn{1}{p{5.4cm}}{Code isolation and sandboxing}& \multicolumn{1}{c}{0.93} & \multicolumn{1}{c}{} & \multicolumn{1}{c}{} & \multicolumn{1}{c}{} & \multicolumn{1}{c}{\checkmark} & \multicolumn{1}{c}{}& \multicolumn{1}{c}{}& \multicolumn{1}{c}{$\bullet$}&  \multicolumn{1}{c}{AV-000}\\
    \multicolumn{1}{p{5.4cm}}{Runtime Application Self-Protection (RASP)}& \multicolumn{1}{c}{0.88} & \multicolumn{1}{c}{} & \multicolumn{1}{c}{} & \multicolumn{1}{c}{\checkmark} & \multicolumn{1}{c}{\checkmark}& \multicolumn{1}{c}{}& \multicolumn{1}{c}{}& \multicolumn{1}{c}{$\bullet$} &  \multicolumn{1}{c}{AV-000}\\
    \multicolumn{1}{p{5.4cm}}{Manual source code review}& \multicolumn{1}{c}{0.85} & \multicolumn{1}{c}{} & \multicolumn{1}{c}{} & \multicolumn{1}{c}{\checkmark} & \multicolumn{1}{c}{}& \multicolumn{1}{c}{}& \multicolumn{1}{c}{$\bullet$}& \multicolumn{1}{c}{$\bullet$}  &  \multicolumn{1}{c}{AV-300}\\ 
     
    \multicolumn{1}{p{5.4cm}}{Build dependencies from sources}& \multicolumn{1}{c}{0.73} & \multicolumn{1}{c}{} & \multicolumn{1}{c}{\checkmark} & \multicolumn{1}{c}{} & \multicolumn{1}{c}{}& \multicolumn{1}{c}{}& \multicolumn{1}{c}{$\bullet$}& \multicolumn{1}{c}{$\bullet$} &  \multicolumn{1}{c}{AV-400, AV-500}\\




    \bottomrule
    \end{tabular}
    }
    \caption{Safeguards against \ac{OSS} supply chain attacks shown in the order of the mean of their Utility-to-Cost ratio assessed by
    17 experts~\cite{ladisa2022taxonomy}. Each safeguard is also characterised by control
    type, stakeholder(s) involved in their implementation, and a mapping to
    mitigated attack vectors (cf. Figure~\ref{fig:attacktreetaxonomy} to resolve their
    identifiers)~\cite{ladisa2022taxonomy}.}
    \label{tab:safeguards}
    \end{table*}

This section presents an overview about safeguards
against \ac{OSS} supply chain attacks, which were identified through literature
review and generalized to become agnostic of specific programming languages or
ecosystems. 


The complete list of the 33 safeguards can be found in Table~\ref{tab:safeguards}, including a classification after
control type and ordered according the Utility-to-Cost ratio as assessed by 17 domain experts~\cite{ladisa2022taxonomy}. 
All safeguards are mapped to the attack vector(s) (described above) they (partially or fully) mitigate, some to
the top-level goal due to addressing all vectors (e.g., establishing
a vetting process), others to more specific subgoals. Some safeguards can be
implemented by one or more stakeholders, while others require the involvement of
multiple ones to be effective (e.g., signature creation and verification).



Both implementation and use of those safeguards can incur non-negligible costs, also depending on the specific programming language and ecosystem. 
Thus, the selection, combination and implementation of safeguards require careful planning and design, to balance
required security levels and costs.

In our work~\cite{ladisa2022taxonomy} we present the qualitative assessment of the utility and cost of each safeguard,
conducted by surveying both domain experts and software developers.


\textit{Common Safeguards}
comprises 4 countermeasures that require all stakeholders to become active,
i.e., project maintainers, open-source consumers, and administrators (service
providers). For example, a detailed \ac{SBOM} has to be produced and maintained
by the project maintainer, 
ideally using automated \ac{SCA} tools. Following, the \ac{SBOM} must be securely hosted and
distributed by package repositories, and carefully checked by downstream users
in regards to their security, quality, and license requirements.

\textit{Safeguards for Project Maintainers and Administrators}
comprises 8 safeguards. \textit{Secure authentication},
for instance, suggests service providers to offer \ac{MFA} or enforce strong
password policies, while project maintainers should follow authentication
best-practices, e.g., use \ac{MFA} where available, avoid password reuse, or
protect sensitive tokens.

\textit{Safeguards for Project Maintainers}
includes 7 countermeasures. Generally, \ac{OSS} projects use
hosted, publicly accessible \ac{VCS}s. Maintainers should then, e.g., conduct
careful \textit{merge request reviews} or enable \textit{branch protection
rules} for sensitive project branches to avoid malicious code
contributions. 
As project builds may still happen on maintainers' workstations, they are
advised to use \textit{dedicated build services}, especially \textit{ephemeral
environments}. 
Additionally, they may \textit{isolate build steps} 
such that they cannot tamper with the output of other build steps.


\textit{Safeguards for Administrators and Consumers} comprises
5 countermeasures. For example, both package repository administrators
and consumers can opt for \textit{building packages directly from source
code}, 
rather than accepting pre-built artifacts. If implemented by package repositories, this would reduce
the risk of subverted project builds. If implemented by consumers, this would
eliminate all risks related to the compromise of 3rd-party build services and
package repositories, as they are taken out of the picture.

\textit{Safeguards for Consumers}
includes 9 countermeasures that may be employed by the
downstream users. The consumers of \ac{OSS} packages may reduce the impact of
malicious code execution when consuming by \textit{isolating the code
and/or sandboxing} it. Another example is the \textit{establishment of internal
repository mirrors} 
of vetted components.

\subsection{Remark}
Some of the presented safeguards are specific to selected
package managers, namely \textit{Scoped packages} (Node.js) and \textit{Prevent
script execution} (Python and Node.js). All others are relevant no matter the
ecosystem, however, control implementations and technology choices differ,
e.g., in case of \textit{Application Security Testing}.

\section{Deeper and deeper: Popularity of Attack Vectors.}

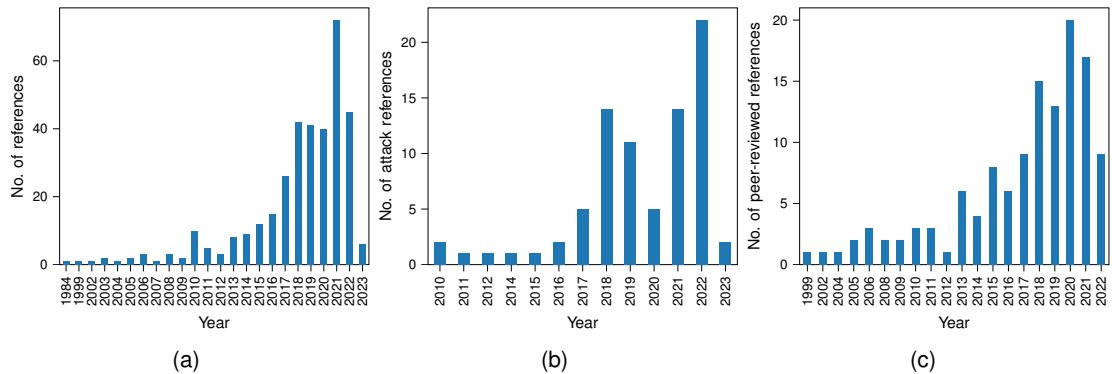
\begin{figure*}[!ht]
    \centering
    \pgfplotsset{compat = 1.3}

\begin{subfigure}[b]{.3\linewidth}
\begin{tikzpicture}[scale=0.6]
  
    \definecolor{darkgray176}{RGB}{176,176,176}
    \definecolor{steelblue31119180}{RGB}{31,119,180}
    
    \begin{axis}[
    tick align=outside,
    tick pos=left,
    xlabel={Year} ,
    ylabel={No. of references},
    x grid style={darkgray176},
    xmin=-0.5, xmax=23.5,
    xtick style={color=black},
    every tick label/.append style={font=\small},
    xtick={0,1,2,3,4,5,6,7,8,9,10,11,12,13,14,15,16,17,18,19,20,21,22,23},
    xticklabel style={rotate=90.0},
    xticklabels={
      1984,
      1999,
      2002,
      2003,
      2004,
      2005,
      2006,
      2007,
      2008,
      2009,
      2010,
      2011,
      2012,
      2013,
      2014,
      2015,
      2016,
      2017,
      2018,
      2019,
      2020,
      2021,
      2022,
      2023
    },
    y grid style={darkgray176},
    ymin=0, ymax=75.6,
    ytick style={color=black}
    ]
    \draw[draw=none,fill=steelblue31119180] (axis cs:-0.25,0) rectangle (axis cs:0.25,1);
    
    \draw[draw=none,fill=steelblue31119180] (axis cs:0.75,0) rectangle (axis cs:1.25,1);
    \draw[draw=none,fill=steelblue31119180] (axis cs:1.75,0) rectangle (axis cs:2.25,1);
    \draw[draw=none,fill=steelblue31119180] (axis cs:2.75,0) rectangle (axis cs:3.25,2);
    \draw[draw=none,fill=steelblue31119180] (axis cs:3.75,0) rectangle (axis cs:4.25,1);
    \draw[draw=none,fill=steelblue31119180] (axis cs:4.75,0) rectangle (axis cs:5.25,2);
    \draw[draw=none,fill=steelblue31119180] (axis cs:5.75,0) rectangle (axis cs:6.25,3);
    \draw[draw=none,fill=steelblue31119180] (axis cs:6.75,0) rectangle (axis cs:7.25,1);
    \draw[draw=none,fill=steelblue31119180] (axis cs:7.75,0) rectangle (axis cs:8.25,3);
    \draw[draw=none,fill=steelblue31119180] (axis cs:8.75,0) rectangle (axis cs:9.25,2);
    \draw[draw=none,fill=steelblue31119180] (axis cs:9.75,0) rectangle (axis cs:10.25,10);
    \draw[draw=none,fill=steelblue31119180] (axis cs:10.75,0) rectangle (axis cs:11.25,5);
    \draw[draw=none,fill=steelblue31119180] (axis cs:11.75,0) rectangle (axis cs:12.25,3);
    \draw[draw=none,fill=steelblue31119180] (axis cs:12.75,0) rectangle (axis cs:13.25,8);
    \draw[draw=none,fill=steelblue31119180] (axis cs:13.75,0) rectangle (axis cs:14.25,9);
    \draw[draw=none,fill=steelblue31119180] (axis cs:14.75,0) rectangle (axis cs:15.25,12);
    \draw[draw=none,fill=steelblue31119180] (axis cs:15.75,0) rectangle (axis cs:16.25,15);
    \draw[draw=none,fill=steelblue31119180] (axis cs:16.75,0) rectangle (axis cs:17.25,26);
    \draw[draw=none,fill=steelblue31119180] (axis cs:17.75,0) rectangle (axis cs:18.25,42);
    \draw[draw=none,fill=steelblue31119180] (axis cs:18.75,0) rectangle (axis cs:19.25,41);
    \draw[draw=none,fill=steelblue31119180] (axis cs:19.75,0) rectangle (axis cs:20.25,40);
    \draw[draw=none,fill=steelblue31119180] (axis cs:20.75,0) rectangle (axis cs:21.25,72);
    \draw[draw=none,fill=steelblue31119180] (axis cs:21.75,0) rectangle (axis cs:22.25,45);
    \draw[draw=none,fill=steelblue31119180] (axis cs:22.75,0) rectangle (axis cs:23.25,6);
    \end{axis}
   
    \end{tikzpicture}
    \subcaption{}
    \label{fig:refs-years-all}
\end{subfigure}
%
%
%
\begin{subfigure}[b]{.3\linewidth}
    \begin{tikzpicture}[scale=0.6]

        \definecolor{darkgray176}{RGB}{176,176,176}
        \definecolor{steelblue31119180}{RGB}{31,119,180}
        
        \begin{axis}[
        tick align=outside,
        tick pos=left,
        xlabel={Year} ,
        ylabel={No. of attack references},
        x grid style={darkgray176},
        xmin=-0.5, xmax=12.5,
        xtick style={color=black},
        xtick={0,1,2,3,4,5,6,7,8,9,10,11,12},
        xticklabel style={rotate=90.0},
        xticklabels={
          2010,
          2011,
          2012,
          2014,
          2015,
          2016,
          2017,
          2018,
          2019,
          2020,
          2021,
          2022,
          2023
        },
        y grid style={darkgray176},
        ymin=0, ymax=23.1,
        every tick label/.append style={font=\small},
        ytick style={color=black}
        ]
        \draw[draw=none,fill=steelblue31119180] (axis cs:-0.25,0) rectangle (axis cs:0.25,2);
        
        \draw[draw=none,fill=steelblue31119180] (axis cs:0.75,0) rectangle (axis cs:1.25,1);
        \draw[draw=none,fill=steelblue31119180] (axis cs:1.75,0) rectangle (axis cs:2.25,1);
        \draw[draw=none,fill=steelblue31119180] (axis cs:2.75,0) rectangle (axis cs:3.25,1);
        \draw[draw=none,fill=steelblue31119180] (axis cs:3.75,0) rectangle (axis cs:4.25,1);
        \draw[draw=none,fill=steelblue31119180] (axis cs:4.75,0) rectangle (axis cs:5.25,2);
        \draw[draw=none,fill=steelblue31119180] (axis cs:5.75,0) rectangle (axis cs:6.25,5);
        \draw[draw=none,fill=steelblue31119180] (axis cs:6.75,0) rectangle (axis cs:7.25,14);
        \draw[draw=none,fill=steelblue31119180] (axis cs:7.75,0) rectangle (axis cs:8.25,11);
        \draw[draw=none,fill=steelblue31119180] (axis cs:8.75,0) rectangle (axis cs:9.25,5);
        \draw[draw=none,fill=steelblue31119180] (axis cs:9.75,0) rectangle (axis cs:10.25,14);
        \draw[draw=none,fill=steelblue31119180] (axis cs:10.75,0) rectangle (axis cs:11.25,22);
        \draw[draw=none,fill=steelblue31119180] (axis cs:11.75,0) rectangle (axis cs:12.25,2);
        \end{axis}
        
        \end{tikzpicture}
        \subcaption{}
        \label{fig:refs-years-attacks}
    \end{subfigure}
%
%
%
\begin{subfigure}[b]{.3\linewidth}
    \begin{tikzpicture}[scale=0.6]

        \definecolor{darkgray176}{RGB}{176,176,176}
        \definecolor{steelblue31119180}{RGB}{31,119,180}
        
        \begin{axis}[
            tick align=outside,
            tick pos=left,
            x grid style={darkgray176},
            xmin=-0.5, xmax=19.5,
            xlabel={Year} ,
            ylabel={No. of peer-reviewed references},
            every tick label/.append style={font=\small},
            xtick style={color=black},
            xtick={0,1,2,3,4,5,6,7,8,9,10,11,12,13,14,15,16,17,18,19},
            xticklabel style={rotate=90.0},
            xticklabels={
              1999,
              2002,
              2004,
              2005,
              2006,
              2008,
              2009,
              2010,
              2011,
              2012,
              2013,
              2014,
              2015,
              2016,
              2017,
              2018,
              2019,
              2020,
              2021,
              2022
            },
            y grid style={darkgray176},
            ymin=0, ymax=21,
            ytick style={color=black}
            ]
            \draw[draw=none,fill=steelblue31119180] (axis cs:-0.25,0) rectangle (axis cs:0.25,1);
            
            \draw[draw=none,fill=steelblue31119180] (axis cs:0.75,0) rectangle (axis cs:1.25,1);
            \draw[draw=none,fill=steelblue31119180] (axis cs:1.75,0) rectangle (axis cs:2.25,1);
            \draw[draw=none,fill=steelblue31119180] (axis cs:2.75,0) rectangle (axis cs:3.25,2);
            \draw[draw=none,fill=steelblue31119180] (axis cs:3.75,0) rectangle (axis cs:4.25,3);
            \draw[draw=none,fill=steelblue31119180] (axis cs:4.75,0) rectangle (axis cs:5.25,2);
            \draw[draw=none,fill=steelblue31119180] (axis cs:5.75,0) rectangle (axis cs:6.25,2);
            \draw[draw=none,fill=steelblue31119180] (axis cs:6.75,0) rectangle (axis cs:7.25,3);
            \draw[draw=none,fill=steelblue31119180] (axis cs:7.75,0) rectangle (axis cs:8.25,3);
            \draw[draw=none,fill=steelblue31119180] (axis cs:8.75,0) rectangle (axis cs:9.25,1);
            \draw[draw=none,fill=steelblue31119180] (axis cs:9.75,0) rectangle (axis cs:10.25,6);
            \draw[draw=none,fill=steelblue31119180] (axis cs:10.75,0) rectangle (axis cs:11.25,4);
            \draw[draw=none,fill=steelblue31119180] (axis cs:11.75,0) rectangle (axis cs:12.25,8);
            \draw[draw=none,fill=steelblue31119180] (axis cs:12.75,0) rectangle (axis cs:13.25,6);
            \draw[draw=none,fill=steelblue31119180] (axis cs:13.75,0) rectangle (axis cs:14.25,9);
            \draw[draw=none,fill=steelblue31119180] (axis cs:14.75,0) rectangle (axis cs:15.25,15);
            \draw[draw=none,fill=steelblue31119180] (axis cs:15.75,0) rectangle (axis cs:16.25,13);
            \draw[draw=none,fill=steelblue31119180] (axis cs:16.75,0) rectangle (axis cs:17.25,20);
            \draw[draw=none,fill=steelblue31119180] (axis cs:17.75,0) rectangle (axis cs:18.25,17);
            \draw[draw=none,fill=steelblue31119180] (axis cs:18.75,0) rectangle (axis cs:19.25,9);
            \end{axis}
        \end{tikzpicture}
        \subcaption{}
        \label{fig:refs-years-peer-review}
    \end{subfigure}
    \caption{Number of collected references per years of publication. (a) References of all types; (b) References discussing real-world attacks; (c) Peer-reviewed references.}
    \label{fig:refs-years}

\end{figure*}

At the time of writing (February 2023) we collected a total of 369 references, of which 81 discuss real-world attacks and 126 are peer-reviewed papers. Figure~\ref{fig:refs-years} shows the number of publications by year for all the references (Fig.~\ref{fig:refs-years-all}), for the references related to real-world attacks (Fig.~\ref{fig:refs-years-attacks}), and for the peer-reviewed references (Fig.~\ref{fig:refs-years-peer-review}).

By grouping the peer-reviewed and real-world attacks references per attack vector we observe the following.

Among the peer-reviewed references we have, in order of popularity: 
\begin{enumerate}
    \item 35 resources discussing the general problem of \textit{Conducting \ac{OSS} Supply Chain Attack (AV-000)};
    \item 33 resources discussing security aspects and issues about \textit{distribution platforms (AV-500)};
    \item 24 resources discussing security aspects and issues of \textit{build systems (AV-400)};
    \item 19 resources discussing security aspects and issues of \textit{\ac{VCS} (AV-300)};
    \item 8 resources discussing problems on packages \textit{creating name confusion with legitimate packages (AV-200)};
    \item 3 resources discussing problems on \textit{malicious packages developed and advertised from scratch (AV-100)}.
\end{enumerate}

For what concerns real-world attacks we have, in order of popularity:
\begin{enumerate}
    \item 26 attacks that exploited the \textit{creation of name confusion with legitimate packages (AV-200)};
    \item 25 attacks that exploited attack vectors in the context of \textit{distribution platforms (AV-500)};
    \item 20 attacks that exploited attack vectors in the context of \textit{\ac{VCS} (AV-300)};
    \item 14 attacks that exploited attack vectors in the context of \textit{build systems (AV-400)};
    \item 7 attacks that consisted of the \textit{development and advertisement of a malicious package from scratch (AV-100)}.
\end{enumerate}

It is straight-forward to observe that the most exploited vector among real-world attacks (i.e., \textit{create name confusion with legitimate packages}) is the least discussed in academic papers. For other attack vectors there is almost a match between attack vectors addressed in peer-reviewed papers and popularity of attacks. The prevention of package name squatting is complex and there exist legitimate uses for which organizations do upload packages with similar names, e.g., using the same prefix for all packages they develop to make them easily recognizable. 
Academic work has begun to propose techniques for the detection of malicious code in package repositories. Though they do not target solely name squatting, one of the safeguards most discussed among peer-reviewed papers, \textit{\ac{AST}} (both static and dynamic), is protecting against such attacks as well as it has the advantage of being general (regardless of whether the malicious package name is squatted or not). However, vetting entire package repositories is computationally burdensome and the false-positive rate must be low so that analyst manual review is practically feasible. 

Another aspect to highlight is the fact that academic works discuss the general problem of \ac{OSS} supply chain attacks and, after \ac{AST}, the most discussed countermeasures are respectively \ac{SCA}/\ac{SBOM} and vulnerability assessment.

\begin{quote}
  The most exploited vector among real-world attacks (i.e., \textit{create name confusion with legitimate packages}) is the least discussed in academic papers.
\end{quote}

\section{A tool to the rescue: SAP's Risk Explorer for Software Supply Chain}

\begin{figure*}[!ht]
	\centering
	\includegraphics[width=1\textwidth]{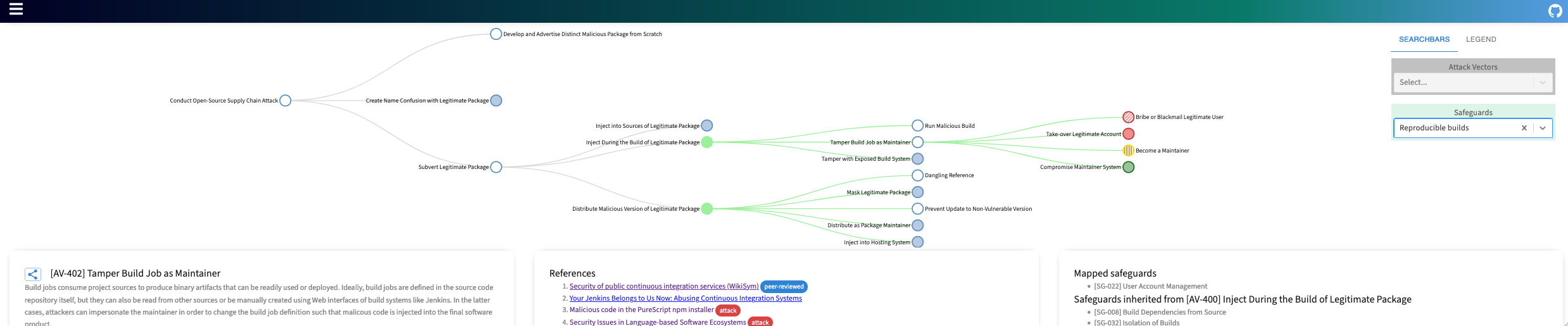}
	\caption{Screenshot of the attack tree visualization (in green attack vectors covered by safeguard \emph{Reproducible builds})~\cite{10.1145/3560835.3564546}.}
	\label{fig:sgview}
  \end{figure*}

  Given the size of the taxonomy and the amount of information associated with it, we develop a companion tool to ease the access and consumption of the taxonomy.
  The tool is open-source\footnote{\url{https://github.com/SAP/risk-explorer-for-software-supply-chains}} to foster the creation of a community that can benefit of and contribute to the taxonomy. When pull requests are merged in the main branch, a new version of the tool is automatically deployed.
  %


  Users can collapse and expand the different nodes of the attack tree 
  to explore the attack surface of open-source-based software development. The 
  description of the respective attack vector, references, as well as associated 
  safeguards are shown below the tree (cf. Figure~\ref{fig:sgview}). 
  This exploratory mode of visualization offers to the user the benefit of managing visual complexity and accessing information in more consumable portions on-demand.
  
  A share button generates a deep link to individual attack vectors, which can be
  referenced from 3rd party websites, e.g., in training material or in security
  advisories that need to reference the attack vector(s) used in a given attack and have a clearer explaination of it.
  A search field in the top-right corner allows searching for attack vectors by
  name. Upon selection, the respective path from the root node to the selected
  attack vector is expanded and highlighted in red.
  The second search field right below is for safeguards. Upon selection, all the
  nodes mitigated by the respective safeguard are colored in green (cf.
  Figure~\ref{fig:sgview}).
  All attack vectors, safeguards, and bibliographical references can
  also be displayed in tabular form. The tabular display of
  attack vectors also allows showing information about associated safeguards in a
  modal window.
  References can be sorted after title, publication year, and affected ecosystem.

\subsection{Industrial Use Cases}\label{sec:usecases}
The Risk Explorer tool can support the following industrial activities.

\textbf{Training and Awareness.} 
The Risk Explorer tool enables an interactive visualization of the taxonomy, the description of attack vectors and safeguards, and the dataset of resources reviewed during the systematization of knowledge. 
Thus, it can be used to better understand security risks in software supply chains and raise awareness among developers and security practitioners.

\begin{figure*}[!htbp]
	\centering
	\includegraphics[width=1\textwidth]{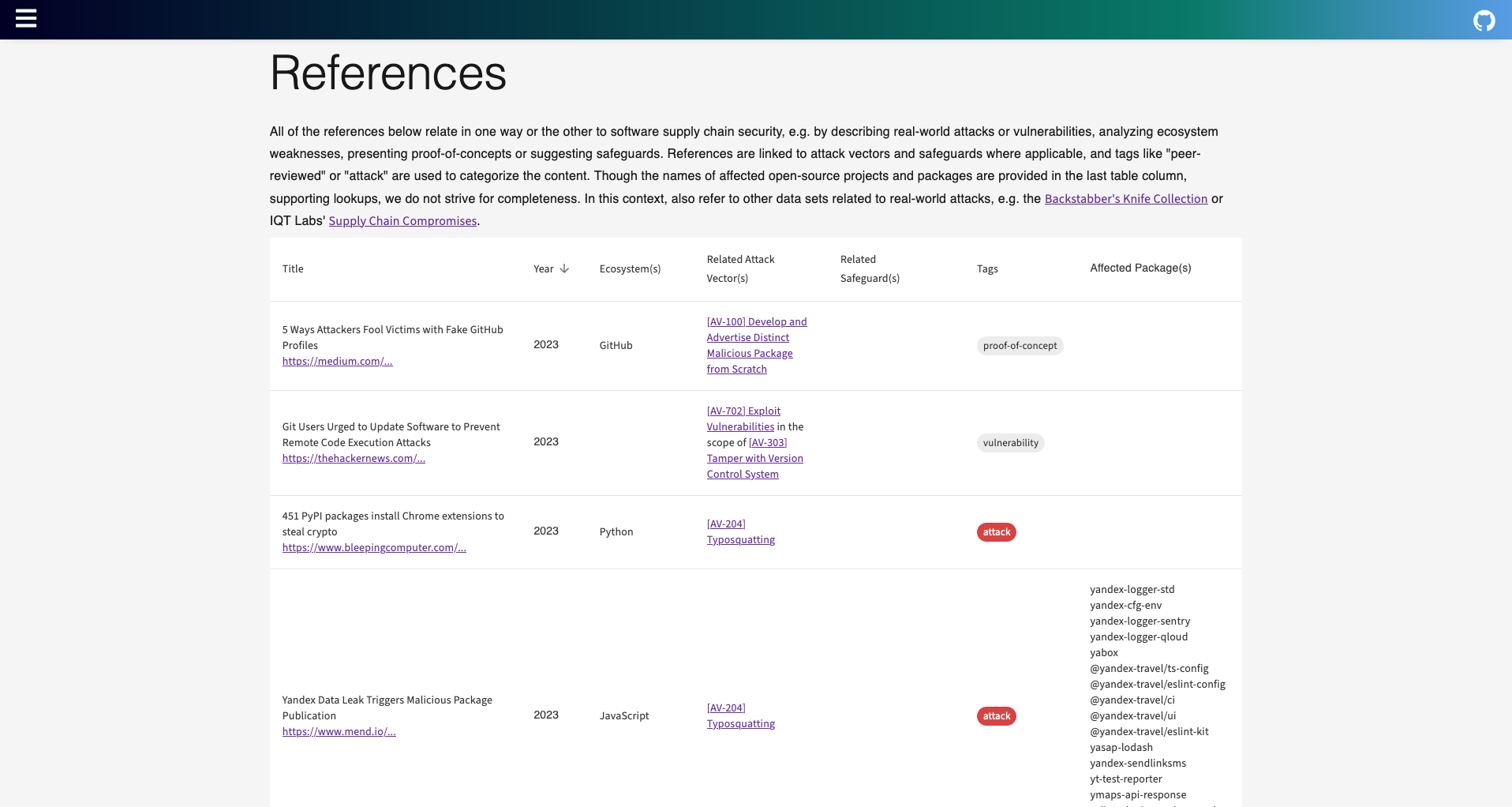}
	\caption{Page containing the tabular view of all the references categorized. They can be sorted by title, year of publication, related attack vectors, related safeguards, and by type.}
	\label{fig:ref-table}
  \end{figure*}

For example, let us assume that we heard about the attack to the \textit{event-stream} package. To figure out which attack vector was used, we can search for such package in the Risk Explorer through the tabular view of references (cf., Fig.~\ref{fig:ref-table}). In the column \textit{Related Attack vectors} we find which attack vector is assigned to. We can then click on the deep link to locate such attack within the taxonomy. At this point, the risk explorer provides the description of the specific attack vector used to infect the \textit{event-stream} package, other attacks that used the same strategy, and mitigating safeguards that would protect against this attack.
Looking at the path from the specific attack vector up to the root node, we can learn about the chain of goals that led to the supply chain attack. By stopping at a certain depth of the taxonomy we can learn about what other possibilities attackers could have used and more generic safeguards that would protect against a broader spectrum of attacks. 

A security expert at SAP used the tool when providing guidance on supply chain
security across the entire organization. She reported that, through the suggested safeguards and
linked references, the tool \emph{"helped learning more about preventive and
detective measures"}. Content, but especially the combination of features
offered by the tool, e.g. \emph{"the visualization and mapping, [and the]
ability to see cross-references"}, was reported as a strength. The expert concluded that
she \emph{"will continue to reference to this research while improving security
practices for development"}.

\begin{quote}
  "[The Risk Explorer] helped learning more about preventive and
  detective measures."
\end{quote}

\textbf{Threat Modeling.} During this activity, the goal is to outline the threats to which a system is (potentially) subjected and then identify possible countermeasures that can be put in place to protect against those threats.

By presenting an extensive list of attack vectors, the tool can support threat modeling activities in the context of \ac{OSS} supply chain attacks. In particular, once the architectural diagram of the infrastructure is outlined, one can identify to which threats (i.e., attack vectors) the actors and systems may be subject to. Moreover, the cross-reference to the related countermeasures helps in understanding how to protect the infrastructure from those threats. 


\textbf{Scope red-team activities.} Red-teaming activities consist of simulating real-world attack scenarios to evaluate the security capability of a system.

Similarly to the ATT\&CK Navigator\footnote{\url{https://mitre-attack.github.io/attack-navigator/}}, the tool can also be used for red/blue team activities planning and security assessments. The enumeration of attack vectors (child nodes) according to different goals (parent nodes) provides a check-list of possible strategies to be adopted (as an attacker) during the security assessment of a project.

\textbf{Gap analysis of safeguards.}
The visual highlight of attack vectors covered by given safeguards through the safeguards searchbar (cf. Fig.~\ref{fig:sgview}) can help in the gap-analysis, 
thus to evaluate which attack vectors are being mitigated by the safeguards in place and which, if any, would need to be implemented.



\subsection{Adding a new attack reference: the case of PyTorch-nightly.}

\begin{figure}[!htbp]
	\centering
	\includegraphics[width=.5\textwidth]{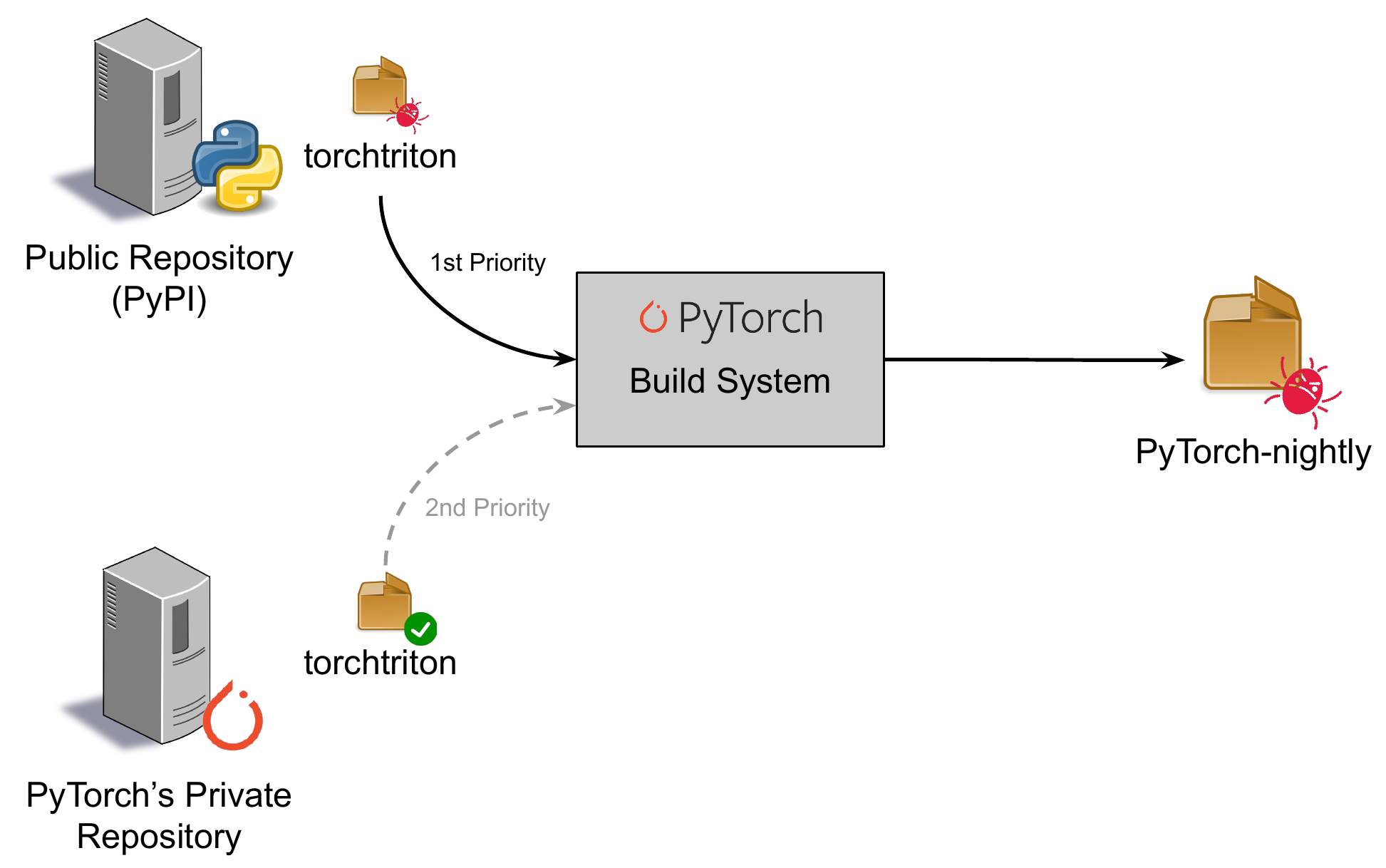}
	\caption{Highlight of the attack vector used in the PyTorch-nightly compromise within the taxonomy.}
	\label{fig:pytorch-attack2}
  \end{figure}

Let us assume that we learn about a new attack and we want to contribute by assigning it to the appropriate attack vector within the tree. We consider as an example the recent PyTorch-nightly's compromise (December 25th-30th, 2022).

The security advisory states that \textit{"a malicious dependency package (torchtriton) [...] was uploaded to the Python Package Index (PyPI) code repository with the same package name as the one we ship on the PyTorch nightly package index. Since the PyPI index takes precedence, this malicious package was being installed instead of the version from our official repository"}~\cite{pytorchPyTorch}. The attack is depicted in Figure~\ref{fig:pytorch-attack2}.

\begin{figure*}[!htbp]
	\centering
	\includegraphics[width=1\textwidth]{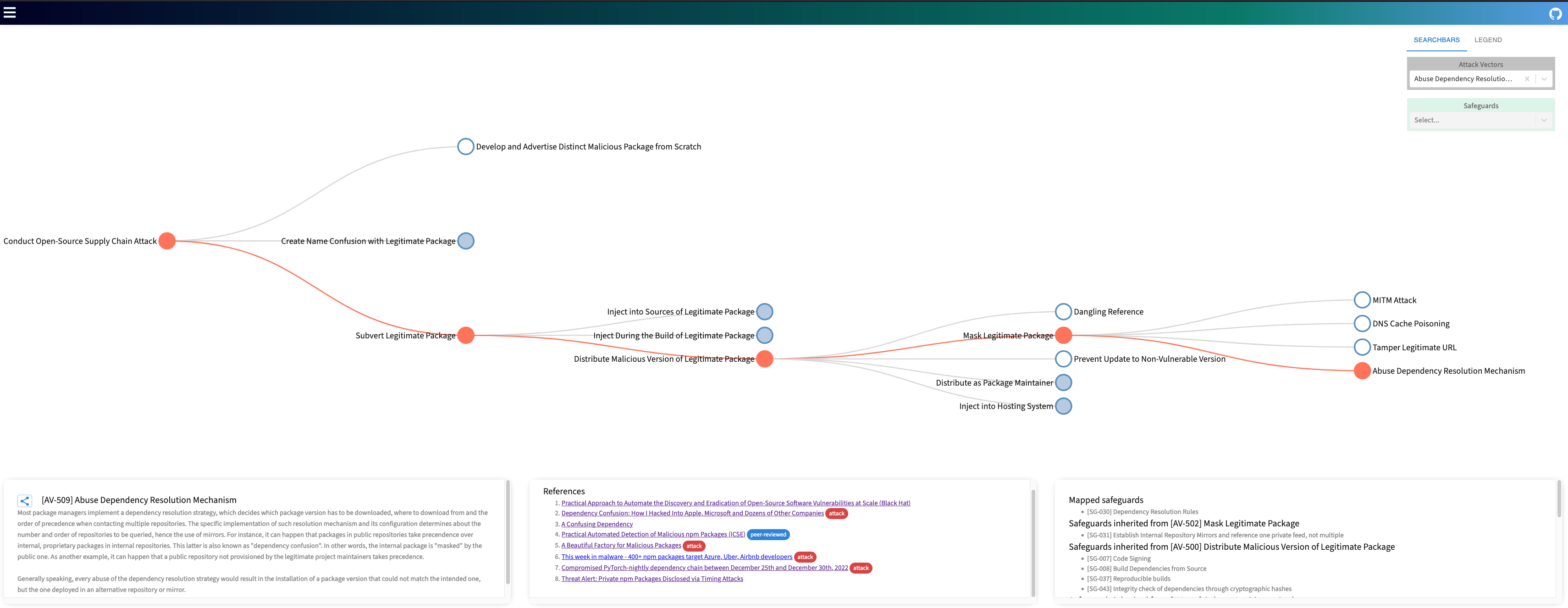}
	\caption{Highlight of the attack vector used in the PyTorch-nightly compromise within the taxonomy.}
	\label{fig:pytorch-attack}
  \end{figure*}

To assign the attack to a node of the taxonomy, we need to navigate it from the root node, establishing which path to follow when moving to deeper nodes. 
From the root node, the first question we ask ourselves is whether the attack consists of developing a malicious package from scratch, exploiting a name similar to that of a legitimate project, or whether a legitimate package has been subverted. Since \texttt{torchtriton} is a legitimate dependency used by PyTorch-nightly, we fall in the last case. From the node \textit{Subvert Legitimate Package} (AV-001), we need to discriminate whether the attack occurred via the \ac{VCS}, build system or the distribution platform. In the compromise of \texttt{torchtriton}, neither the source code nor the build infrastructure were compromised, the attack occurred in the PyPI package repository and thus we fall in the last node \textit{Distribute Malicious Version of Legitimate Package} (AV-500). By reading the security advisory, the root cause of the attack was caused by the fact that \texttt{torchtriton} was only present in the PyTorch nightly package index (used to build the PyTorch-nightly project) and not in the public PyPI repository. Thus, the legitimate dependency \texttt{torchtriton} was \textit{masked} by \textit{abusing the dependency resolution mechanism} (AV-509). The PyTorch-nightly's compromise can thus be assigned to node (AV-509) (cf. Fig.~\ref{fig:pytorch-attack}). 

Whenever an attack exploits new ways of injecting malicious code, it may happen that the existing vectors may not covering the newly identified behavior. In such cases an extension of the taxonomy itself has to be considered.

\section{We're not alone: Related frameworks}


Open-source and other organizations have recognized the importance of securing open-source software and open-source based supply chains in the development of secure software and services. 
Several frameworks have been proposed to provide practices for a secure software lifecycle.

The Enduring Security Framework\footnote{\url{https://www.nsa.gov/About/Cybersecurity-Collaboration-Center/Cybersecurity-Partnerships/ESF/}} (a public-private cross-sectional working group led by NSA, CISA, and ODNI) released actionable guidance to developers, suppliers, and customers to secure the entire software supply chain.
“Securing the Software Supply Chain for Developers” targets the software development lifecycle and considers threats to the development of secure code, to the verification of third-party components, to the hardening of the build system, and to the delivering of code and provides recommended mitigations.
“Securing the Software Supply Chain for Suppliers” focuses on how vendors should identify threats that could compromise their organization, software development, software product, and software delivery.
“Securing the Software Supply Chain for Customers” focuses on the best practices for the acquisition, deploymnet, and operation of the software product.


The Microsoft \ac{OSS} \ac{S2C2F}\footnote{\url{https://www.microsoft.com/en-us/securityengineering/opensource/osssscframeworkguide}} combines requirements and tools to reduce risks associated with the consumption of open-source software. It is based on three core concepts: control all consumed open-source software, use of maturity model to help in prioritizing the requirements to implement, and secure the software supply chain at scale. Compared to the framework “Securing the Software Supply Chain for Developers”, the Microsoft \ac{S2C2F} only focuses on the secure consumption of open source software, but provides a more detailed guidance in this context. It defines 8 practices (e.g., scan and update third-party components) and a list of associated operational requirements (e.g., scan OSS for malware, perform security reviews of OSS). It also comes with a maturity model that organizes the requirements into 4 different levels and lists tools that can support fulfilling the requirements. 

The OWASP \ac{SCVS}\footnote{\url{https://owasp.org/www-project-software-component-verification-standard/}} aims at establishing a framework for identifying activities, controls, and best practices, which can help in identifying and reducing risk in a software supply chain. Similarly to \ac{S2C2F}, the OWASP \ac{SCVS} provides 6 families of controls (e.g., inventory, pedigree and provenance) and 3 levels of verification requiring an increasing number of requirements for higher assurance. However, it applies to software components in general and thus the requirements are not specific for OSS consumed within the supply chain like in the case of the \ac{S2C2F} framework. 

The \ac{SLSA}\footnote{\url{https://slsa.dev/}} is an OpenSSF project and consists of a checklist of standards and controls to prevent tampering, improve integrity, and secure packages and infrastructure in your projects, businesses or enterprises. It defines four levels of assurance, from simple provenance information via a documented, automated build process, to high confidence and trust via peer-review of source code changes with hermetic, reproducible builds. Compared to Microsoft \ac{S2C2F} and OWASP \ac{SCVS}, \ac{SLSA} has a narrower scope as it focuses on integrity, thus ensuring that the consumed code has not been tampered. 


The different frameworks overlap as they all aim to provide guidance and recommendations and some efforts to map them have been done, e.g., the OSS SSC framework includes a mapping of requirements to other relevant specifications including OWASP \ac{SCVS} and \ac{SLSA}. However, all frameworks lack a systematic representation of how attacks can be carried out in order to clearly define what are the threats covered by the provided recommentations. 

The taxonomy presented in this work complements such efforts. It takes the perspective of the attacker and provides the most complete taxonomy of attacks whereas the frameworks above focus on safeguards by providing recommendations. Similar to what the MITRE ATT\&CK framework does for attacks to infrastructures, it helps in describing how attackers can exploit the software supply chain to spread malwares. Having a complete taxonomy is key for establishing whether the identified recommendations cover the known attack vectors. By providing a base of knowledge related to attacks, it helps in designing and researching novel countermeasures and may also facilitate the comparison of the existing frameworks.

\section{From attack vectors to malicious code}

As mentioned before, the scope of the taxonomy is to describe \textit{how} attackers can inject malicious code in upstream projects, regardless the actual malicious content injected.
To complement, looking at the malicious code provides valuable information about the behavior the attackers are trying to achieve and how different ecosystems are affected.

\paragraph{Malware Behavior}
As first mentioned in~\cite{ohm2020backstabbers}, the main objectives observed in existing \ac{OSS} supply chain attacks are:

\begin{itemize}
    \item \textit{Reverse shell}, which consists in spawning a shell process and redirects both its input and output through an open socket to the attacker machine.
    \item \textit{Droppers} connect to an attacker-controlled host to download a second-stage payload that will be then executed. The remote payload can be read directly through the connection or be temporarily stored in a local file.
    \item \textit{Data exfiltration} (most common behavior~\cite{ohm2020backstabbers}) aims at reading sensitive information (e.g., environment variables, files) to then sends it to an attacker-controlled endpoint.
    \item \textit{\ac{DoS}} is typically achieved either through resource exhaustion (e.g., fork-bombs) or by deleting system files.
    \item \textit{Financial gain} obtained by executing crypto miners in the target system. 
\end{itemize}

During a recent analysis we performed on the NPM and PyPI ecosystem\footnote{Paper under submission to a conference that requires anonymity}, we also observed the presence of \textit{research proofs-of-concept} aiming at testing execution functionalities offered by the language and/or the specific package manager to show the potential risks when installing such packages. Another unwanted behavior we observed is the presence of \textit{rickrolling attacks}. When we observed and reported such packages, the resepective security teams decided not to remove them as they do not consider rickrolling as malicious behavior.

\paragraph{Malware Techniques}
We also observed cases where obfuscation techniques are used. 
%
Also note that code obfuscation may be more or less interesting for an attacker depending on the ecosystem.
In case of interpreted languages, downloaded packages contain the malware's source
code, which makes it more accessible to analysts compared to compiled languages. 
The presence of encoded or encrypted code in such packages prooved being a good
indicator of compromise~\cite{sejfia2022practical}, as there are few legimitate
use-cases for open-source packages (e.g., minification, mostly used for frontend JavaScript libraries). Still, the quantity of open-source packages and versions makes
manual inspection very difficult, even if source code is accessible.
\cite{ohm2020backstabbers} showed that malicious packages usually have no obfuscation or use simple techniques (e.g., base64), however a small fraction of the packages we studied used obfuscation with more complex mechanisms, such as encrypting the code (e.g., AES-256) or with custom encodings of strings and identifiers.
When it comes to compiled code, well-known techniques like packing, dead-code
insertion or subroutine reordering 
make reverse engineering and analysis more complex. 

For an attack to succeed, malicious code needs to be executed. Attackers achieve this either at installation time, during test-cases, or at runtime (e.g., by
embedding the payload in a specific function or initializer).
The execution at installation time is possible in only certain ecosystems.
For Python and Node.js, this is commonly achieved through installation hooks, which trigger
the execution of code provided in the downloaded package (e.g., in
\texttt{setup.py} for Python or \texttt{package.json} for JavaScript). A
comparable feature is not present in most compiled languages, like Java or C/C++. 

To increase the chances of succeeding, attackers may conduct malware campaigns, which consist in the distribution of a large number of packages that implement the same malicious behavior to increase the reach towards downstream users.
These campaigns can affect only one ecosystem or be cross-language (i.e., same malicious behavior implemented in different languages and spread to related ecosystems).

\paragraph{Malware Reporting}
How malicious packages are reported to the respective security teams varies from one ecosystem to another. For example, NPM offers the reporting functionality only through UI within the official website: the reporter has to search for the package in a searchbar, navigate to the project page, click on the button 'Report Malware', and finally fill in the report form. If multiple reports are submitted in a short time frame, CAPTCHA checks are enabled in the website. For PyPI, instead, the official procedure to report malicious packages consists in sending an e-mail to the security team\footnote{\url{https://pypi.org/security/}} with the names of the packages and (preferably) the link to the lines of code containing the malware highlighted using \textit{Inspector}\footnote{\url{https://inspector.pypi.io}}.
Becaue of the way they are structured, these procedures make it complex to report multiple malwares at the same time, especially considering the popularity of malware campaigns.

\begin{quote}
    "[current reporting mechanisms] make it complex to report multiple malwares at the same time, especially considering the popularity of malware campaigns."
  \end{quote}

  Once reported and confirmed, malicious packages are removed from package repositories and are no longer publicly accessible. Given the significance of the problem and the interest from both academia and industry in developing detection methods for software supply chain attacks, it would be beneficial to maintain a dataset of these attacks (both packages and associated metadata).

  \begin{quote}
    "Once reported and confirmed, malicious packages are removed from package repositories and are no longer publicly accessible. [...] it would be beneficial to maintain a dataset of these attacks (both packages and associated metadata)."
  \end{quote}





\section{The journey ended? Benefits and open challenges.}

Our work systematizes knowledge about \ac{OSS} supply chain security by
abstracting, contextualizing and classifying existing works. The proposed
taxonomy can benefit future research by offering a central point of reference
and a common terminology.
The comprehensive list of attack vectors and safeguards can support assessing
the security level of open-source projects, e.g., to conduct comparative empiric
studies across projects and ecosystems and over time.

An open challenge in OSS supply chain attacks is the detection of malicious code. 
The availability of source code in ecosystems for interpreted languages suggests
that malware analysis is more straight-forward. Still, recent publications focus
on those ecosystems, especially JavaScript and
Python~\cite{ladisa2022taxonomy},
partly due to their popularity, but also because existing malware analysis
techniques cannot be easily applied.
More subtle attacks, such as intentional insertion of vulnerabilities, complicate detection since they 
require analysis of the context of the change to distinguish it from an accidentally introduced vulnerability. 
Additionally, code generation and the difficulty in identifying \ac{VCS} commits that correspond to pre-built components
make malware analyis of source code difficult.


\bibliographystyle{plain}

\bibliography{bibliography.bib}

\begin{acronym}[TDMA]
    \acro{CTI}{Cyber Threat Intelligence}
    \acro{TUF}{The Update Framework}
    \acro{PKI}{Public Key Infrastructure}
    \acro{CI}{Continuous Integration}
    \acro{CD}{Continuous Delivery}
    \acro{UI}{User Interface}
    \acro{VCS}{Versioning Control System}
    \acro{VMs}{Virtual Machines}
    \acro{VA}{Vulnerability Assessment}
    \acro{VCS}{Version Control System}
    \acro{SCM}{Source Control Management}
    \acro{IAM}{Identity Access Management}
    \acro{CDN}{Content Delivery Network}
    \acro{UX}{User eXperience}
    \acro{SLR}{Systematic Literature Review}
    \acro{SE}{Social Engineering}
    \acro{MITM}{Man-In-The-Middle}
    \acro{SBOM}{Software Bill of Materials}
    \acro{MFA}{Multi-Factor Authentication}
    \acro{AST}{Application Security Testing}
    \acro{RASP}{Runtime Application Self-Protection}
    \acro{OSS}{Open-Source Software}
    \acro{TARA}{Threat Assessment and Remediation Analysis}
    \acro{CAPEC}{Common Attack Pattern Enumeration and Classification}
    \acro{DoS}{Denial of Service}
    \acro{SCA}{Software Composition Analysis}
    \acro{SLSA}{Supply-chain Levels for Software Artifacts}
    \acro{SDLC}{Software Development Life-Cycle}
    \acro{ICT}{Information and Communication Technologies}
    \acro{C-SCRM}{Cyber Supply Chain Risk Management}
    \acro{DDC}{Diverse Double-Compiling}
    \acro{OSINT}{Open Source Intelligence}
    \acro{U/C}{Utility-to-Cost}
    \acro{PII}{Personally Identifiable Information}
	\acro{CWE}{Common Weakness Enumeration}
  \acro{S2C2F}{Secure Supply Chain Consumption Framework}
  \acro{SCVS}{Software Component Verification Standard}
  \acro{SLSA}{Supply chain Levels for Software Artifacts}
\end{acronym}

\end{document}